\documentclass[aps,amsmath,amssymb,twocolumn,prb,longbibliography]{revtex4-2}
\usepackage{graphicx} 
\usepackage{amsfonts,amsmath,amssymb}
\usepackage{amsthm}
\usepackage{dcolumn}
\usepackage{dsfont,bm}
\usepackage[colorlinks=true,linkcolor=blue,pagecolor=blue,filecolor=blue,menucolor=blue,urlcolor=blue,citecolor=blue,anchorcolor=blue]{hyperref}
\usepackage{xcolor}
\usepackage{soul} 
\usepackage{amsbsy}
\usepackage{float}

\usepackage{bm}

\begin{document}

	\title{Magnon Polarization Textures and Transport in Antiferromagnets and Altermagnets}

	\author{Konstantin S. Denisov}	
	\email{denisokonstantin@gmail.com}
	\author{Igor {\v{Z}}uti{\'c}}
	
	\affiliation{Department of Physics, University at Buffalo, State University of New York, Buffalo, NY 14260, USA}

	\begin{abstract} 

Antiferromagnets (AFMs) host magnons that can exhibit fully circular, elliptic and linear polarizations of sublattice magnetization dynamics. We analyze theoretically different momentum-space magnon polarization textures and magnon transport in collinear AFMs and altermagnets (AMs). The sensitivity of the polarization texture to a magnon momentum direction determines two regimes of magnon polarization kinetics: The Dyakonov-Perel relaxation or its dephasing due to asynchronous Larmor precession. 
For an easy-axis AFM, magnetic field parallel to the N{\'e}el vector leads to the suppression of the magnetic dipolar interaction-induced Dyakonov-Perel relaxation, accompanied by the increase in the magnon spin-diffusion length and its counter-intuitive growth with temperature. For a perpendicular orientation, instead, relaxation of circular polarization of magnons is enhanced due to Larmor precession. Finally, extremely anisotropic relaxation of magnon polarization is predicted for AMs or AFMs having Dzyaloshinskii-Morya interaction.

\end{abstract}

	\date{\today}
	\maketitle

Magnons, propagating waves of magnetic moment oscillation carrying integer angular momentum, 
are elementary quantum excitations of magnetically-ordered matter~\cite{bloch1930theorie,kittel1963quantum}. 
Faster and more efficient methods of storing and processing information include magnon-based spintronics~\cite{kruglyak2010magnonics,lenk2011building,pirro2021advances,flebus2024roadmap}, which exploits the functionalities of spin and angular momenta degree of freedom~\cite{vzutic2004spintronics,Tsymbal:2019}. 
Driven in magnetic insulators, magnons address many of these challenges by offering the THz range dynamics and low energy dissipation due to absence of electrical current, 
which can be further adjusted for classical~\cite{chumak2014magnon,chumak2022advances,flebus2024roadmap} or quantum~\cite{lachance2019hybrid,li2020hybrid,serha2026magnetic} computations.

Magnon phenomena are particularly interesting in antiferromagnets  (AFMs)~\cite{baltz2018antiferromagnetic,hortensius2021coherent,han2023coherent,rimmler2025non}, 
magnetically-ordered systems with zero magnetization but finite sublattice-resolved staggered magnetic moments. 
In AFMs, there is at least one pair of inequivalent magnetic moments in a unit cell, 
signalling a multiband character of magnonic states. 
In a collinear AFM, 
two different magnon species carry positive and negative angular momenta, $\pm \hbar$, 
and its mixed state features elliptic polarizarion of magnetization dynamics. 
Multiple magnon flavours enables 
a conversion between different  polarizations~\cite{shiota2024handedness,sun2024dipolar,leenders2024canted},  
spin-momentum locking~\cite{okuma2017magnon,kawano2019designing,kawano2019discovering,shen2020magnon},  spin relaxation~\cite{shen2020magnon,hu2026anisotropic,gao2026observation}, 
non-Abelian braiding~\cite{daniels2018nonabelian}, or 
the Hanle effect~\cite{wimmer2020observation,kamra2020antiferromagnetic,guckelhorn2023observation,kleinherbers2024magnon}.  

Significance of magnon polarizations is especially important when a finite magnon energy splitting is present. This can be due to relativistic Dzyaloshinskii-Morya (DM) interaction, 
also enabling nonreciprocal
magnon transport~\cite{iguchi2015nonreciprocal}. 
In AFMs with magnetic sublattices not connected by spatial translations and inversion symmetry~\cite{pekar1964combined,vsmejkal2020crystal,yuan2020giant,hayami2020bottom,yuan2021prediction}, 
also referred to as altermagnets (AMs)~\cite{krempasky2024altermagnetic,song2025altermagnets,duan2025antiferroelectric,cao2025symmetry,zhu2026altermagnetic,jungwirth2026symmetry}, 
the spin group symmetry analysis~\cite{brinkman1966theory,litvin1974spin,liu2022spin,chen2024enumeration,liu2026symmetry}  
has identified another fundamental type of spin splittings due to a nonrelativistic exchange interaction.  
Similar to the altermagnetic splitting of electronic bands~\cite{krempasky2024altermagnetic,song2025altermagnets,duan2025antiferroelectric,cao2025symmetry,zhu2026altermagnetic,jungwirth2026symmetry}, a counterpart mechanism is discussed for magnons~\cite{cui2023efficient,chen2025unconventional,sodequist2024two,liu2024chiral,yu2024chiral,gomonay2024structure,cui2025anisotropic,sun2025observation,sears2026altermagnetic}. 

Mixing between AFM magnons, which determines the magnetization dynamics polarization, 
and their population imbalance 
can be described by a pseudospin, as in Stokes parameters for the polarization and intensity of light~\cite{jackson1962classical}. In addition to magnon density and its current, a more complete description involves transport and dynamics of magnon polarization, encoded by its pseudospin~\cite{wimmer2020observation,kamra2020antiferromagnetic,guckelhorn2023observation,kleinherbers2024magnon}. 
Exploiting this extra degree of freedom could expand magnon-based applications, 
but its verification is still incomplete. 
Promising techniques include 
the quantum impurity sensing~\cite{flebus2018quantum,wang2022noninvasive,heitzer2024characterization,melendez2025quantum,bittencourt2026quantum}, magnon injection~\cite{lebrun2018tunable,ross2019propagation,troncoso2020spin,wimmer2020observation,van2024magnon}, spatially resolved Brillouin light scattering~\cite{olsson2020pure,kargar2021advances}, X-rays~\cite{gu2025observing,wittrock2026soft}, or Sagnac interferometry~\cite{karimeddiny2023sagnac}. 
These advances call for theoretical 
understanding of possible magnon polarization transport regimes in emerging AFMs with magnon energy splittings.

\begin{figure*}[t]
	\centering
	\includegraphics[width=.16\textwidth]{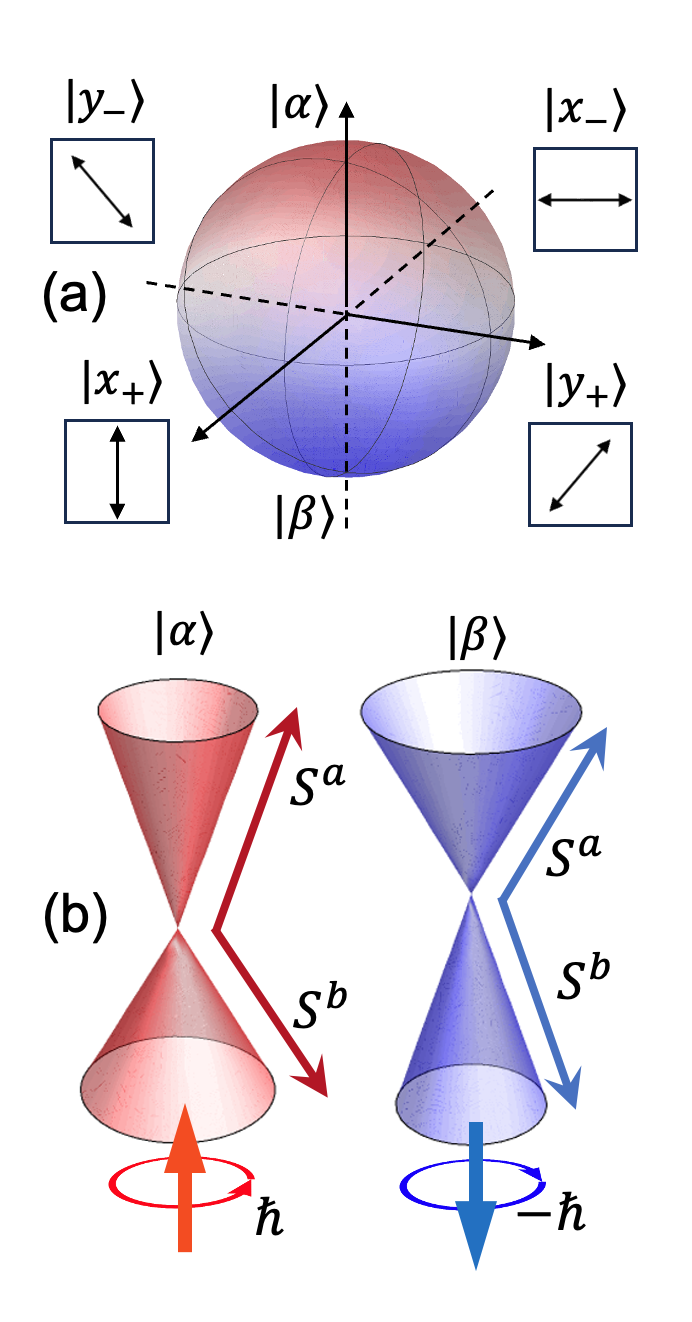}
	\includegraphics[width=.55\textwidth]{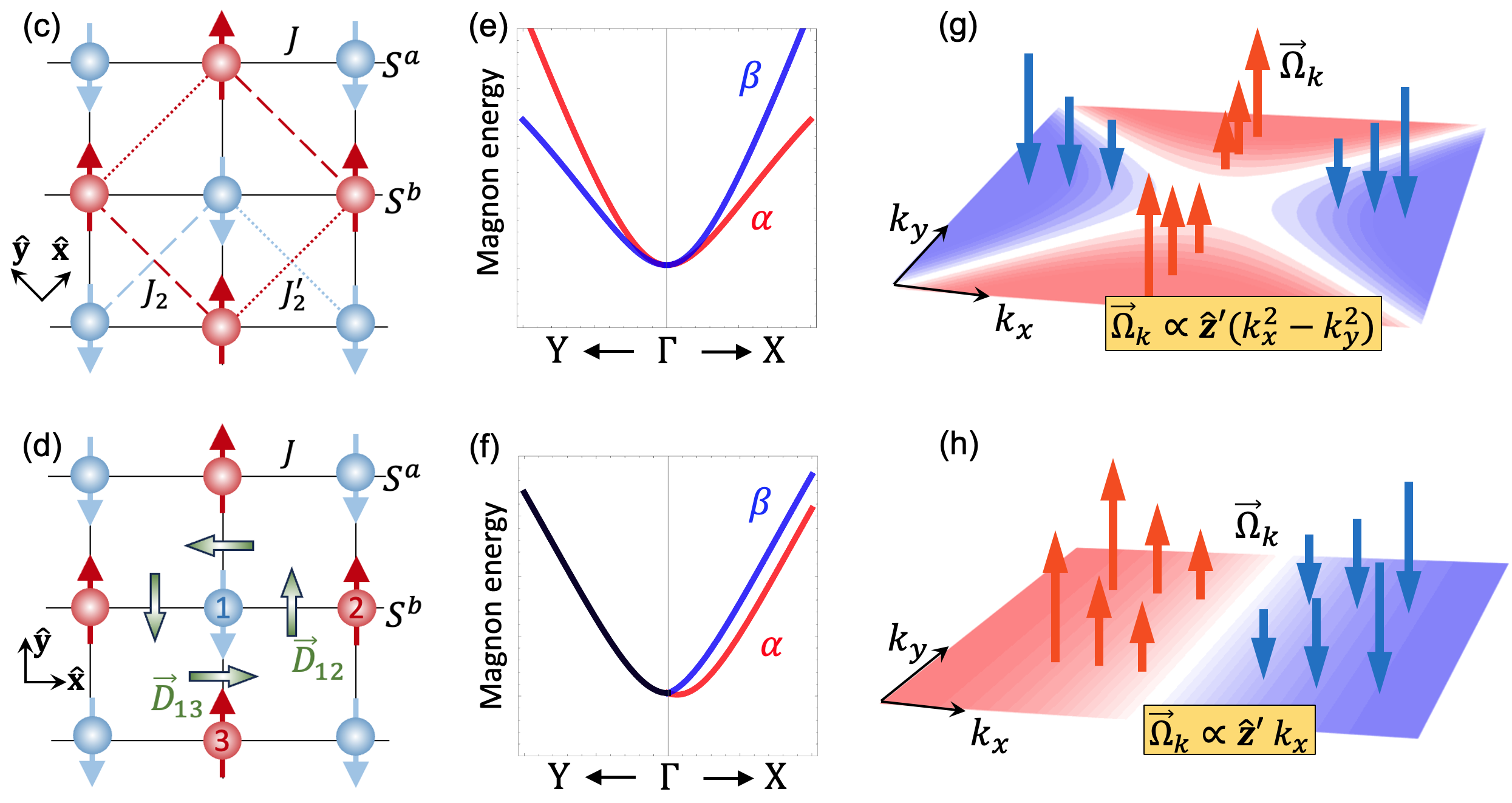}
	\includegraphics[width=.17\textwidth]{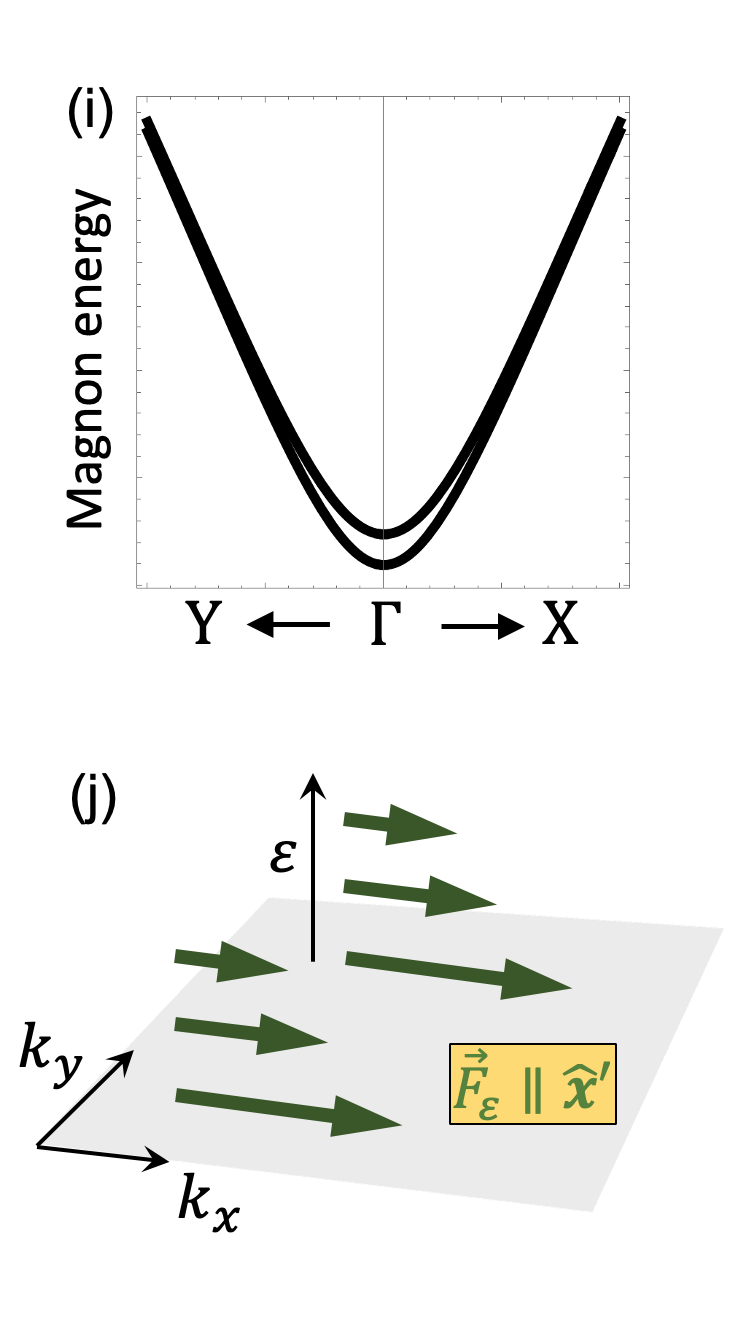}
	\caption{(a) Bloch sphere for the magnon pseudospin 
		and the orientation of linear polarizations at its equator.
		(b) Circularly polarized dynamics of spins in $a,b$-sublattices  
		for $\alpha$ and $\beta$ magnons.  
		(c) 2D AM Lieb lattice. Dashed and dotted lines depicting the strength of the second-nearest neighbor exchange interactions, $J_2 \neq J_2'$. 
		(d) 2D square AFM lattice. 
		The green arrow pointing in the direction of the DM interaction, $\bm{D}_{ij}= \hat{\bm{z}} \times \bm{R}_{ij}$. 
		(e),(f) Magnon band stucture and 
		(g),(h) the distribution of momentum-dependent Larmor frequency vector $\bm{\Omega}_{\bm k}$ in the Brillouin zone for AFM lattices from (c),(d). 
		(i),(j) Magnon dispersion and its Larmor frequency vector $\bm{F}_\varepsilon$ distribution for the canted phase of an easy-axis AFM with perpendicular $\bm{B}_0$.} 
	\label{fig:Magnons}
\end{figure*}

In this work we address this issue by considering dynamics and transport of diffusive magnon polarization in collinear AFMs and AMs.  We examine possible magnon polarization textures and discuss the realization of two different polarization dynamics regimes. (i) When a magnon energy splitting varies with its momentum direction there is a motional narrowing effect and its Dyakonov-Perel (DP) relaxation~\cite{dyakonov1972spin} 
of elliptic magnon polarization~\cite{shen2020magnon,hu2026anisotropic,gao2026observation}, enhanced significantly in AMs and AFMs with a finite DM interaction. (ii) A dephasing of magnon  polarization~\cite{wimmer2020observation,kamra2020antiferromagnetic,guckelhorn2023observation,kleinherbers2024magnon} 
due to asynchronous precession with energy-dependent Larmor frequency, 
present due to a magnetic anisotropy or an external field, ${\bm B}_0$.   
We use the Boltzmann kinetic theory and the magnon density matrix to analyze these mechanisms, their interplay and manifestation in the magnon nonlocal transport  accessible in spin-valve structures with magnon injection.

A common AFM with two magnetic sublattices has the following magnetic Hamiltonian
\begin{multline}
	\mathcal{H}_0 = \sum_{\langle  i,j \rangle } J_{ij} \bm{S}_i^a \bm{S}_j^b
+ \sum_{i,j} K \left[  (S_{i,z}^{a})^2  + (S_{j,z}^{b})^2  \right] + 
\\
  + \hbar \gamma (\bm{S}_i^a + \bm{S}_j^b) \bm{B}_0 + \mathcal{H}_{\rm DM} + \mathcal{H}_{\rm dip}, 
\end{multline}
here $\bm{S}_{i}^{a}, \bm{S}_{j}^{b}$, are the spin operators for $a,b$-sublattices at sites $i,j$, 
the first term is the exchange interaction ($J_{ij}>0$) with the summing, $ \langle i,j  \rangle $, over $i,j$ lattice sites, 
the second term is the magnetic anisotropy with positive (negative) $K$ for the easy-plane (easy-axis) anisotropy, 
 $\gamma>0$ is the gyromagnetic ratio, 
 and $\mathcal{H}_{\rm {DM, dip}}$ stands for DM and magnetic dipolar interactions, respectively.

Magnon energy band splitting can be induced by different lattice structures, see Fig.~\ref{fig:Magnons} and  Supplemental Material (SM)~\cite{SM}.  
We consider (i) An altermagnetic structure based on a two-dimensional (2D) Lieb lattice with its 
orientation-dependent second-nearest neighbor exchange interaction, $J_2 \neq J_2'$~\cite{cui2023efficient}, see Fig.~\ref{fig:Magnons}(c);  (ii) a 2D square lattice lacking horizontal (in-plane) mirror symmetry 
and allowing for a finite DM term, $\bm{D}_{ij} \left( \bm{S}_i^a \times  \bm{S}_j^b \right)$, 
between neighboring sites separated by $\bm{R}_{ij}$, with the DM vector $\bm{D}_{ij}  = D \hat{\bm{z}} \times \bm{R}_{ij}/R_{ij}$, Fig.~\ref{fig:Magnons}(d); (iii) magnetic dipolar interaction, $\mathcal{H}_{\rm dip}$,~\cite{shen2020magnon}.

Magnons are boson excitations above a magnetic ground state. For large $S^{a,b} \approx S$, the latter can be determined classically by the minimization of the magnetic free energy with respect to the orientation of the sublattice magnetizations, $\bm{M}^{a,b}$. In the absence of $B_0$, we deal with a collinear AFM, $\bm{M}^a = - \bm{M}^b$ with the N{\'e}el vector, $\bm{L} = \bm{M}^a - \bm{M}^b$. A finite $\bm{B}_0 \perp \bm{L}$ results in the canting of $\bm{M}^{a,b}$ with $\bm{M}^a + \bm{M}^b \parallel \bm{B}_0$. In this work we consider only small canting angles $\theta_0 \ll 1$, the role of $\theta_0 \sim 1$ for magnon states can be found~\cite{okuma2017magnon,kawano2019discovering,kawano2019thermal}.

To derive the effective Hamiltonian for magnons one employs a standard procedure~\cite{kittel1963quantum} consisting of (i) the Holstein-Primakoff transformation applied to the spin operators, $\bm{\mathcal{S}}_{i,j}^{a,b}$, 
in the rotated basis along~$-\bm{M}^{a,b}$
[the basis vectors set by $-\bm{M}^{a,b}$ 
are denoted $(\hat{\bm{n}}_x, \hat{\bm{n}}_y, \hat{\bm{n}}_z)$]: 
${\mathcal{S}}_{i,z}^a = S - a_i^\dagger a_i $ and 
${\mathcal{S}}_{i,\pm}^a \approx \sqrt{2S} (a_i, a_i^\dagger)$ 
(the same for $\bm{\mathcal{S}}_j^b$ with $b_j, b_j^\dagger$), 
where $a_i,b_j$ ($a_i^\dagger,b_j^\dagger$) are the Bose creation (annihilation) operators, 
(ii) applying the rotation matrices, $\bm{S}_{i,j}^{a,b} = U^{a,b}(\theta_0) \bm{\mathcal{S}}_{i,j}^{a,b}$,
and (iii) taking the Fourier components in the momentum space, $a_{\bm{k}}, b_{\bm{k}}$. 
The resulting Hamiltonian, $\mathcal{H}$, 
can be presented as a block matrix 
in the basis 
$\Psi_{\bm k}^T = (a_{\bm{k}}, b_{-\bm{k}}^\dagger, b_{\bm{k}}, a_{-\bm{k}}^\dagger)$
\begin{equation}
	\label{eq:Hblock}
	\mathcal{E} = \sum_{\bm k>0} \Psi_{\bm k}^\dagger \mathcal{H} \Psi_{\bm k}, 
	\qquad
		\mathcal{H} = \begin{pmatrix}
		\mathcal{H}_1^+ & \mathcal{H}_2 \\
		\mathcal{H}_2^\dagger & \mathcal{H}_1^-
	\end{pmatrix}, 
\end{equation}
where $\mathcal{H}_{1,2}$ are $2 \times 2$ matrices specified below. 
To obtain the magnon energies one should diagonalize an axillary matrix $g \mathcal{H}$, 
where $g = {\rm diag}(1,-1,1,-1)$ to comply with bosonic commutation rules for the Bogoliubov transformation~\cite{white1965diagonalization,colpa1978diagonalization,rezende2019introduction,kawano2019discovering,kawano2019thermal}. 
However, a full matrix structure of $\mathcal{H}$ hides a couple of independent effects and, for analyzing the magnon polarization textures, it is more instructive to proceed in a different way~\cite{SM}.

Magnons in AFM with a well-defined angular momentum projection along $\bm{L}$ are set by $\mathcal{H}_1^\pm$. 
For a collinear AFM, $\mathcal{H}_1^\pm$ can be expressed as 
\begin{equation}\label{eq:H1}
	\mathcal{H}_1^\pm =  \begin{pmatrix}
		\mathcal{A}_{\bm{k}}  & - \mathcal{B}_{\bm{k}}^\pm \\
		- \mathcal{B}_{\bm{k}}^\pm & \mathcal{A}_{\bm{k}} 
	\end{pmatrix}
\mp \begin{pmatrix}
	\Delta_{\bm k} & 0 \\
	0 & - \Delta_{\bm k}
\end{pmatrix}, 
\end{equation}
where $\mathcal{A}_{\bm k}, \mathcal{B}_{\bm k}^\pm, \Delta_{\bm k}$ 
are functions of $\bm{k}$ determined by an AFM lattice~\cite{SM}. 
To diagonalize $\mathcal{H}_1^\pm$, we apply the Bogoliubov transformation with 
$\alpha$ and $\beta$ magnon operators 
$\beta_{\bm{k}} = u_{\bm{k}} a_{\bm{k}} - v_{\bm{k}} b_{-{\bm{k}}}^\dagger$, 
$\alpha_{-{\bm{k}}}^\dagger = v_{\bm{k}} a_{\bm{k}} - u _{\bm{k}} b_{-{\bm{k}}}^\dagger$, 
and $u_{\bm{k}}^2-v_{\bm{k}}^2=1$ for Bose statistics. 
The energies are
\begin{equation}\label{eq:Bogolub}
	\varepsilon_{\bm{k}}^{\alpha,\beta} = \varepsilon_{\bm{k}}^\mp \pm \Delta_{\bm k}, 
	\qquad
	\varepsilon_{\bm{k}}^\pm  = \sqrt{\mathcal{A}_{\bm{k}}^2 - (\mathcal{B}_{\bm{k}}^\pm)^2}, 
\end{equation}
with the coefficients $(u_{\bm{k}},v_{\bm{k}}) = \sqrt{{(\mathcal{A}_{\bm{k}} \pm \varepsilon_{\bm{k}}^\pm)}/{2 \varepsilon_{\bm{k}}^\pm }}$. 
For $\theta_0 \ll 1$, $\alpha$ and $\beta$ magnons carry angular momentum, $\pm \hbar$
\begin{equation}\label{eq:SpinZ}
	\mathcal{S}_z = \sum_{\bm{k}}  - a_{\bm{k}}^\dagger a_{\bm{k}} + b_{\bm{k}}^\dagger b_{\bm{k}}  = \sum_{\bm{k}} \alpha_{\bm{k}}^\dagger \alpha_{\bm{k}} - \beta_{\bm{k}}^\dagger \beta_{\bm{k}}. 
\end{equation}

A linear combination of $\alpha_{\bm k}$ and $\beta_{\bm k}$ 
describes a magnon state with an elliptic polarization of sublattice magnetization dynamics. 
We can treat the doublet of states ($\alpha_{\bm k}, \beta_{\bm k}$) 
with the pseudospin and its Pauli matrices vector, $\bm{\sigma}$~\cite{shen2020magnon,nambu2020observation,kamra2020antiferromagnetic,wimmer2020observation}; 
the basis vectors in the pseudospin space are $(\hat{\bm{x}}', \hat{\bm{y}}', \hat{\bm{z}}')$. 
Figure~\ref{fig:Magnons}(a) shows the Bloch sphere for $\bm{\sigma}$ and the associated magnon mode polarizations. 
The north and south poles are the two eigenstates of $\sigma_z$ corresponding to $\alpha$ and $\beta$ magnons 
with counter-clockwise ($\hbar$) and clockwise ($-\hbar$) rotations of sublattice spins $\pm \hat{\bm{n}}_z$, respectively, see Fig.~\ref{fig:Magnons}(b). The eigenstates of $\sigma_x$ and $\sigma_y$, denoted as $|x_\pm \rangle$ and $|y_\pm \rangle$, correspond to almost linearly polarized spin oscillations along $\hat{\bm{n}}_x$ and $\hat{\bm{n}}_y$ for $|x_\pm \rangle$, and shifted by $45^\circ$ for $|y_\pm \rangle$.

A dependence of $\bm{\sigma}$ on the magnon momentum, $\bm{k}$, 
defines a magnon pseudospin texture, $\bm{\sigma}_{\bm k}$. In contrast to electronic states, 
which exhibit a great variety of spin configurations sensitive to a particular point-group symmetry of a material~\cite{dyakonov2017spin}, there is only a limited number of possible magnon pseudospin textures attainable for a collinear AFM~\cite{cheng2016antiferromagnetic,daniels2018nonabelian,kawano2019designing,kawano2019discovering}.  
Having in mind the role of $\bm{\sigma}_{\bm k}$ for the magnon polarization transport, 
we can differentiate them by casting an effective $2\times 2$ 
magnon Hamiltonian in the basis $(\alpha_{\bm{k}}, \beta_{\bm{k}})$
\begin{equation}\label{eq:2by2}
	\mathcal{H}_{\bm{k}} = \varepsilon_{\bm{k}} + 
	\hbar (\bm{\Omega}_{\bm{k}} + \bm{F} )  \bm{\sigma}, 
\end{equation}
where we separated ${\bm k}$-orientation dependent and independent 
Larmor vectors, $\bm{\Omega}_{\bm k}$ and $\bm{F}$,  respectively, different examples are shown in Fig.~\ref{fig:Magnons}; 
here $\varepsilon_{\bm{k}} = (\varepsilon_{\bm{k}}^+ + \varepsilon_{\bm{k}}^- )/ 2$.

First, there is a collinear $\bm{\sigma}_{\bm k} \parallel \hat{\bm{z}}'$ 
due to a finite energy splitting between $\alpha_{\bm k}$ and $\beta_{\bm k}$  
given by $\hbar \Omega_{\bm k}$; typically it increases away from the $\Gamma$ point. 
For 2D AMs~\cite{SM}, $\hbar {\Omega}_{\bm k} = \Delta_{\bm k} \propto (J_2 - J_2') (k_x^2 - k_y^2)$, at small $\bm{k}$ it has a signature $d$-wave dependence. 
For a 2D square lattice with a finite $\mathcal{H}_{\rm DM}$ and $\bm{L} \parallel \hat{\bm{y}}$~\cite{SM}, we have ${\Omega}_{\bm k} = \gamma(\mathcal{B}_{\bm k}^+ - \mathcal{B}_{\bm k}^-) \propto D f(\varepsilon) k_x $, with $f(\varepsilon)$ a function of energy, $\varepsilon$. The distribution of $\bm{\Omega}_{\bm k}$ in the Brouillon zone (BZ)
for these two cases is shown in Figs.~\ref{fig:Magnons}(g),(h). 

Second, there is another collinear $\bm{\sigma}_{\bm k} \parallel \hat{\bm{x}}'$, 
driven by $\bm{F} \parallel \hat{\bm{x}}'$ and enforcing linear polarization of  magnetization dynamics $\parallel \hat{\bm{n}}_x$. This situation is realized for the in-plane $K>0$ and $\bm{L} \perp \hat{\bm{z}}$~\cite{kawano2019designing,kleinherbers2024magnon}, 
or due to a canting of $\bm{M}^{a,b}$ induced by $B_0$~\cite{kamra2020antiferromagnetic,wimmer2020observation}, including the easy-axis with $\bm{L} \parallel \hat{\bm{z}}$ and $\bm{B}_0 \perp \hat{\bm{z}}$. 
Despite that $\bm{F}$ is $\bm{k}$-orientation independent, 
it still depends on the magnon energies, $\bm{F}_{\varepsilon} \equiv \bm{F}$, 
decreasing away from $\Gamma$ point, see Fig.~\ref{fig:Magnons}(e).  
To obtain the term $\bm{F} \bm{\sigma}$, we account for $\mathcal{H}_2$ in Eq.~(\ref{eq:Hblock}), 
change the basis from $a_{\bm{k}}, b_{\bm{k}}$ to the magnon operators $\alpha_{\bm{k}}, \beta_{\bm{k}}$, and keep a finite coupling only between $\alpha_{\bm{k}}$ and $\beta_{\bm{k}}$. 
Mixing effects between $\alpha_{\bm{k}}$ and $\alpha_{-{\bm{k}}}^\dagger$ ($\beta_{\bm k}$ and $\beta_{-\bm k}^\dagger$) 
are negligible when $\varepsilon_{k=0}\neq 0$. 
When both $\bm{\Omega}_{\bm k}$ and $\bm{F}$ are present, $\bm{\sigma}_{\bm k}$ rotates with $\bm{k}$ in the $(x'z')$ plane.

{A special case is the magnetic dipolar interaction}~\cite{shen2020magnon}: $\bm{\Omega}_{\bm{k}}  = \Omega_{\bm{k}} 
\sin^2{\theta_k} ( \hat{\bm{x}}' \cos{2 \phi_k} + \hat{\bm{y}}' \sin{2 \phi_k})$, 
where $\theta_k$ and $\phi_k$ are the azimuthal and polar angles of $\bm{k}$, 
$\Omega_{\bm{k}} = \gamma_{\bm{k}} {\omega_A \varepsilon_m}/{2 \varepsilon_{\bm{k}}}$, 
$\hbar \omega_A = 2K S$, and $\varepsilon_m = \hbar \gamma 4 \pi M_s$ with the sublattice-saturation magnetization $M_s$. 
Here, the pseudospin texture exhibits vortex-like rotation in $(x'y')$ plane with $\bm{k}$,
 its magnitude, $\Omega_k \propto \varepsilon^{-1}$, gets weaker for magnons away from the $\Gamma$ point of BZ.

To study the role of different pseudospin textures for magnon transport, 
we make use of the magnon density matrix in the semiclassical approximation
\begin{equation}
	\rho_{\bm{k}}(\bm{r},t) = N_{\bm{k}}(\bm{r},t) + \bm{P}_{\bm{k}}(\bm{r},t) \bm{\sigma}, 
\end{equation}
where $N_{\bm{k}}$ is the average distribution function of $\alpha$ and $\beta$ magnons, 
and $\bm{P}_{\bm{k}}$ is the pseudospin distribution function. 
The components of $\bm{P}_{\bm{k}}$ can be seen as the Stokes parameters, in analogy with light polarizations~\cite{landau1975classical}. 
The kinetic Boltzmann equation for 
$\bm{P}_{\bm k}$ is
\begin{equation}\label{eq:KinBol}
	\frac{\partial \bm{P}_{\bm k}}{\partial t} + (\bm{v}_{\bm k} \nabla ) \bm{P}_{\bm k} - \left[ (\bm{F}  + \bm{\Omega}_{\bm k}) \times \bm{P}_{\bm k} \right] = \bm{G}_{\bm k} + \mathcal{I}[\bm{P}_{\bm k}], 
\end{equation}
here $\bm{v}_{\bm k} = \partial \varepsilon_{\bm k} / \partial \bm{k}$ is the magnon velocity, 
$\bm{G}_{\bm k}$ is the generation term and 
$\mathcal{I}$ is the collision integral modelled~as 
\begin{equation}\label{eq:St}
\mathcal{I}[\bm{P}_{\bm k}]
= - \frac{\bm{P}_{\bm k} - \langle \bm{P}_{\bm k} \rangle}{\tau} - \frac{\bm{P}_{\bm k}-\bm{P}_{k}^0}{\tau_g}. 
\end{equation}
The first term accounts for the elastic magnon momentum relaxation 
with time $\tau$, 
$\langle . \rangle$ stands for $\bm{k}$-direction averaging. 
The second term describes the magnon number nonconserving relaxation~\cite{zhang2012magnon,zhang2012spin} 
towards an equilibrium $\bm{P}_k^0$; $\tau_g$ can be related to the Gilbert damping~\cite{ritzmann2014propagation,cramer2018spin,kleinherbers2024magnon}. 
We are interested in the leading effects 
of low-energy magnon polarization transport. This allows us to  
disregard a possible matrix structure of the velocity operator~\cite{shen2020magnon,kawano2019thermal,kovalev2016spin}  
and to use the isotropic band approximation, $\varepsilon = \varepsilon_{\bm k}$, with constant $v = v_{\bm k}$.

\begin{figure}[t]
	\centering
	\includegraphics[width=.45\textwidth]{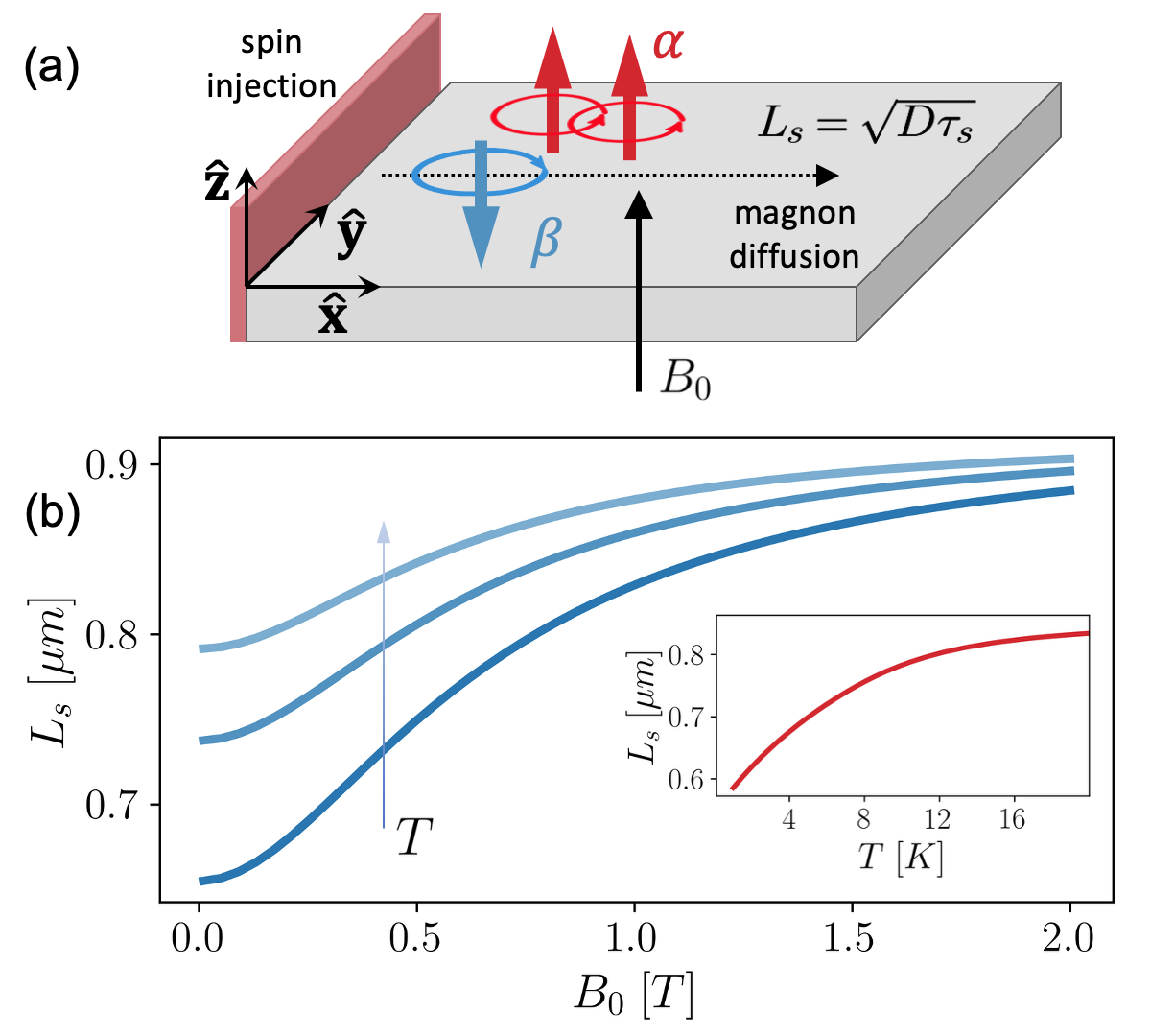}
	\caption{(a) Diffusion of magnons from spin injection region. 
		(b) The increase in $L_s$ 
		with $B_0$ due to 
		dipolar interaction-induced $\Gamma_z(B_0)$ suppression 
		 for different temperatures ($T=10,15,30$~K) for MnF$_2$ and $\tau_{g} = 150$~ns. The inset shows $L_s$ dependence on $T$ at $B_0=0$ for a 
		$T$-independent $\tau$.} 
	\label{fig:s_T_and_B}
\end{figure}

We focus on the diffusion-dominated regime with $\tau$ being the shortest timescale 
($\tau \ll \tau_g$ and $\tau \ll \Omega_{\bm k}^{-1}, F^{-1}$)
and the mean free path, $\ell = v \tau $, small compared to 
spatial gradients of $\rho_{\bm k}(\bm{r},t)$ and all geometrical lengths. 
The dynamics of the pseudospin density 
\begin{equation}
	\bm{P}(\bm{r},t) = \sum_{\bm k} \bm{P}_{\bm k} = \int d\varepsilon \nu_\varepsilon 
	\bm{P}_\varepsilon(\bm{r},t), 
\end{equation}
changes for $t\gg \tau$, $r\gg \ell$. Here $\bm{P}_\varepsilon \equiv  \langle  \bm{P}_{\bm k}  \rangle$, and $\nu_\varepsilon$ is the isotropic density of states. 
The first term in Eq.~(\ref{eq:St}) leaves $\bm{P}_\varepsilon$ unchanged and  
only higher angular harmonics of $\bm{P}_{\bm k}$ decay fast with $\tau$.   
A drift-diffusion equation for $\bm{P}_\varepsilon(\bm{r},t)$ is obtained 
after integrating out the angular-dependent parts of $\bm{P}_{\bm k}$ in Eq.~(\ref{eq:KinBol}) in the lowest order of its coupling to $\bm{P}_\varepsilon$, valid for $\Omega_{\bm k} \tau \ll 1$ 
\begin{equation}\label{eq:DriftDiff}
	\frac{\partial \bm{P}_\varepsilon}{\partial t} - D \nabla^2 \bm{P}_\varepsilon - 
	\left[  \bm{F}_{\varepsilon} \times \bm{P}_\varepsilon \right] +  \left(\hat{\Gamma} +\tau_{g}^{-1}\right) \bm{P}_\varepsilon = 0, 
\end{equation}
where $D = v^2 \tau /d$ is the diffusion coefficient ($d=2,3$)
and the spin injection at the left contact [$x=0$, see Fig.~\ref{fig:s_T_and_B}(a)] is modelled by a boundary condition for the spin current flowing along $\hat{\bm{x}}$: $\bm{J} = - D \bm{P}_\varepsilon'  |_0 \propto \hat{\bm{z}}' G_\varepsilon =  \hat{\bm{z}}' (-\partial N_\varepsilon^0 / \partial \varepsilon)$, here $N_\varepsilon^0$ is the Bose-Einstein equilibrium distribution function of magnons in AFM~\cite{zhang2012magnon,zhang2012spin,kleinherbers2024magnon}.

We note the appearance of an additional relaxation of $\bm{P}_\varepsilon$ described by a tensor, 
$\hat{\Gamma} = {\rm diag}(\Gamma_x, \Gamma_y, \Gamma_z)$, 
\begin{equation}
\label{eq:GammaDP}
	\Gamma_i = \langle  \left(\bm{\Omega}_{\bm k}^2  - \Omega_{{\bm k},i}^2  \right)\tau \rangle. 
\end{equation}
This is a manifestation of the DP spin-relaxation mechanism~\cite{dyakonov1972spin} due to 
the motionial narrowing. 
Physically, it originates from an erratic precession of a magnon pseudospin subjected to an effective magnetic field, $\hbar \bm{\Omega}_{\bm k}/\gamma$, randomly varying in time due to scattering-induced diffusion. The DP relaxation for magnons is essentially anisotropic: For AMs and DM-equipped lattices from Fig.~\ref{fig:Magnons}, 
the collinear $\bm{\Omega}_{\bm{k}} = \Omega_{\bm{k}} \hat{\bm{z}}'$ leads only to the relaxation of the in-plane pseudospin components and its elliptic/linearly polarization ($\Gamma_x = \Gamma_y \neq 0$), 
with no damping ($\Gamma_z=0$) for  the magnon spin density $S_z \equiv P_z$. 
For the magnetic dipolar interaction~\cite{shen2020magnon} 
all three components of $\hat{\Gamma}$ are nonzero with $\Gamma_z = 2 \Gamma_{x,y} =~\langle \bm{\Omega}_k^2 \tau \rangle$.

Separating the full Larmor frequency in two parts, 
$\bm{\Omega}_{\bm k}$ and $\bm{F}_{\varepsilon}$, and considering type (a) and (b) $\bm{\sigma}_{\bm k}$ textures 
is fully justified here based on the pseudospin dynamics perspective of Eq.~(\ref{eq:DriftDiff}): While $\bm{\Omega}_{\bm k}$ leads to the extra relaxation of $\bm{P}_{\varepsilon}$, the vector $\bm{F}_{\varepsilon}$ simply leads to the Larmor precession of $\bm{P}_{\varepsilon}$ around it.

We next analyze 
nonlocal magnon spin transport and its static pseudospin accumulation, $\bm{P}(x)$, 
in a spin-valve geometry at a distance $x$ from the injector, see Fig.~\ref{fig:s_T_and_B}(a). 
At $\bm{F} \parallel \hat{\bm z}'$, 
we have  $P_z$  decoupled from $P_{x,y}$, and $S_z(x)$ is 
\begin{equation}\label{eq:Szx}
	S_z(x) = \int d\varepsilon \nu_\varepsilon P_{\varepsilon,z}^0 e^{ - x / L_\ast(\varepsilon)},
	\quad 	L_\ast(\varepsilon) = \sqrt{D \tau_\ast}, 
\end{equation}
with $P_z^0 \propto (-\partial N_\varepsilon^0 / \partial \varepsilon)$, 
and  $\tau_\ast^{-1} = \Gamma_z + \tau_{g}^{-1}$. 
The resulting integral fits very well by $S_z(x)\propto e^{-x/L_s}$ 
with some effective $L_s$ and spin-relaxation time, $\tau_s = L_s^2/D$, set by the competition between the Gilbert damping with $\tau_g^{-1}$ and the DP relaxation of $P_z$,  described by $\Gamma_z$.

For AFM magnons, $\Gamma_z \neq 0$ is present for the dipolar interaction. 
Interestingly, for an easy-axis configuration with $K<0$ and $\bm{L} \parallel \hat{\bm{z}}$, 
the rate $\Gamma_z$ can be controlled by the application of $\bm{B}_0 \parallel \hat{\bm{z}}$. 
The magnitude of $\Gamma_z$ is renormalized according to (with $\Gamma_{x,y} = \Gamma_z /2 $)
\begin{align}\label{eq:DPRates}
	& 	\Gamma_z(B_0) = \frac{ \langle  (\Omega_{\bm{k},x}^2 + {\Omega}_{\bm{k},y}^2 ) \tau \rangle}{1 + (\gamma B_0 \tau)^2} = \frac{\Gamma_z(B_0=0)}{1 + (\gamma B_0 \tau)^2}.
\end{align}
We obtained $\Gamma_z(B_0)$ by the same procedure of decoupling 
$\bm{P}_\varepsilon$ from finite angular harmonics of $\bm{P}_{\bm k}$ 
by including $\bm{F} = \hat{\bm{z}} \gamma B_0$ in its dynamics. 
The  suppression of $\Gamma_z \propto \Omega_k^2 / (\gamma  B_0)^2 \tau \ll \Omega_k^2 \tau$ at $\gamma B_0 \tau \gg 1$ follows the spin-relaxation suppression for mobile carriers in  semiconductors~\cite{ivchenko1973spin,glazov2004magnetic,wu2010spin,rakitskii2025resonant}
and AMs~\cite{denisov2024anisotropic}.  
Applying $B_0 \hat{\bm{z}}$ removes the DP relaxation of $S_z$ visible in Fig.~\ref{fig:s_T_and_B}(b), where we show the $L_s(B_0)$ in MnF$_2$ for different temperatures, $T$. 
Note the recent observation of the $B_0$-induced magnon spin relaxation suppression~\cite{gao2026observation}. 
Our energy-dependent analysis elaborates on details of the mechanism.

We recall that $\Omega_k \propto \varepsilon^{-1}$, for dipolar interaction, 
so $\Gamma_z \propto \varepsilon^{-2}$, 
(assuming constant $\tau$) and gets weaker for magnons away from $\Gamma$ point. 
At higher $T$, the population of magnons with larger $\varepsilon$ and smaller $\Gamma_{z}$ 
grows, which hinders the DP relaxation and can lead to a counterintuitive increase of $L_s$ with $T$, 
shown in the inset of Fig.~\ref{fig:s_T_and_B}(b). 
Accordingly, the $B_0$-induced suppression of $\Gamma_z$ also weakens at larger $T$, visible in  Fig.~\ref{fig:s_T_and_B}(b).

For estimating  $\Gamma_z$ in real AFMs, first we note that 
a momentum relaxation time of magnons is expected to be much longer than for electrons/holes.  
For $\tau = 15$~ps~\cite{wang2022noninvasive},  
we get $\gamma B_0 \tau \approx 1$ already at $B_0 = 0.4$~T, 
in contrast to at least several tesla for mobile carriers~\cite{ivchenko1973spin,glazov2004magnetic,wu2010spin}. 
Second, as a model example in Fig.~\ref{fig:s_T_and_B}(b)  we consider MnF$_2$, its magnetic parameters are 
$H_E = 57$~T, $H_A=0.58$~T and 
$\varepsilon_{k=0} \approx 1.1$~meV~\cite{rezende2019introduction}; 
here $H_{E} = z JS/\hbar \gamma$ ($z$ is the number of nearest neighbors) and $H_A = 2 KS/\hbar \gamma$. 
With $\varepsilon_m \approx 0.08$~meV
and $\tau =15$~ps ($D= 0.5 \times 10^{-5}$ m$^2$s$^{-1}$,  $v=1$~km/s), 
for magnons at the  bottom of BZ
we get $\Gamma_z^{-1} \approx 4$~ns ($\Omega_{k=0} \approx 6$~GHz); 
$\Omega_{k=0} \tau \approx 0.08$, justifying Eq.~(\ref{eq:GammaDP}). 
For MnF$_2$, the DP-mechanism of $S_z$ relaxation 
is observable when $\tau_{g}$ exceeds tens of ns, likely realizable at low $T$. 
However, the importance of $\Gamma_z$ 
varies across different materials. 
For another conventional AFM, Cr$_2$O$_3$,
with $H_E = 250$~T, $H_A=0.07$~T, $\varepsilon_m = 0.07$~meV~\cite{rezende2019introduction},   
the same estimation yields $\Omega_{k=0} \approx 0.6$~GHz 
and $\Gamma_z^{-1} \approx 300$~ns, making this relaxation less effective.

Including $\bm{F}\parallel \hat{\bm{x}}'$ results in the linear magnon polarization. 
We consider a previously unreported case of 
the canted phase for an easy-axis AFM with 
$\bm{L} \parallel \hat{\bm{z}}$ and $\bm{B}_0 \parallel \hat{\bm{x}}$. 
In equilibrium, $\bm{M}^{a,b} = M (-\sin{\theta_0},0, \mp \cos{\theta_0})$, where 
$\theta_0 \approx  {B_0}/{(H_A+2H_E)}$. 
The derived $\bm{F}_\varepsilon$ is given by
\begin{equation}
	\bm{F}_\varepsilon = \hat{\bm{x}}'  \theta_0^2 
	\gamma_k \gamma H_E
	{\mathcal{A}_k }/{\varepsilon_k}. 
\end{equation}
$F_\varepsilon$ is only slightly anisotropic with respect to $\bm{k}$, 
which we ignore assuming the isotropic band approximation.  The energy-dependence of $\bm{F}_\varepsilon$ 
results in the depolarization of $S_z$. For an isotropic pseudospin relaxation, at time $\tau_g$, the two eigenvalues of static Eq.(\ref{eq:DriftDiff}) are $ \lambda_\pm^2 = {L_\ast^{-2} \pm i {F}_\varepsilon/ D}$, 
with $L_\ast^2 = D \tau_g$. After an initial excitation, $\bm{P}_\varepsilon(t=0) \parallel \hat{\bm{z}}'$, the precession follows $P_{\varepsilon,z}(t) \propto \cos{(F_\varepsilon t)} e^{-t/\tau_g}$ and $P_{\varepsilon,y}(t) \propto \sin{(F_\varepsilon t)} e^{-t/\tau_g}$. For reasonable $T$, the dynamics of overall $S_z(t)$ will accumulate $\bm{P}_\varepsilon(t)$ precessing asynchronously at different $F_\varepsilon$, resulting in the $B_0$-induced damping of $S_z(x)$, also  referred to as the magnon Hanle effect~\cite{kamra2020antiferromagnetic,wimmer2020observation,kleinherbers2024magnon}.

\begin{figure}[t]
	\centering
	\includegraphics[width=.4\textwidth]{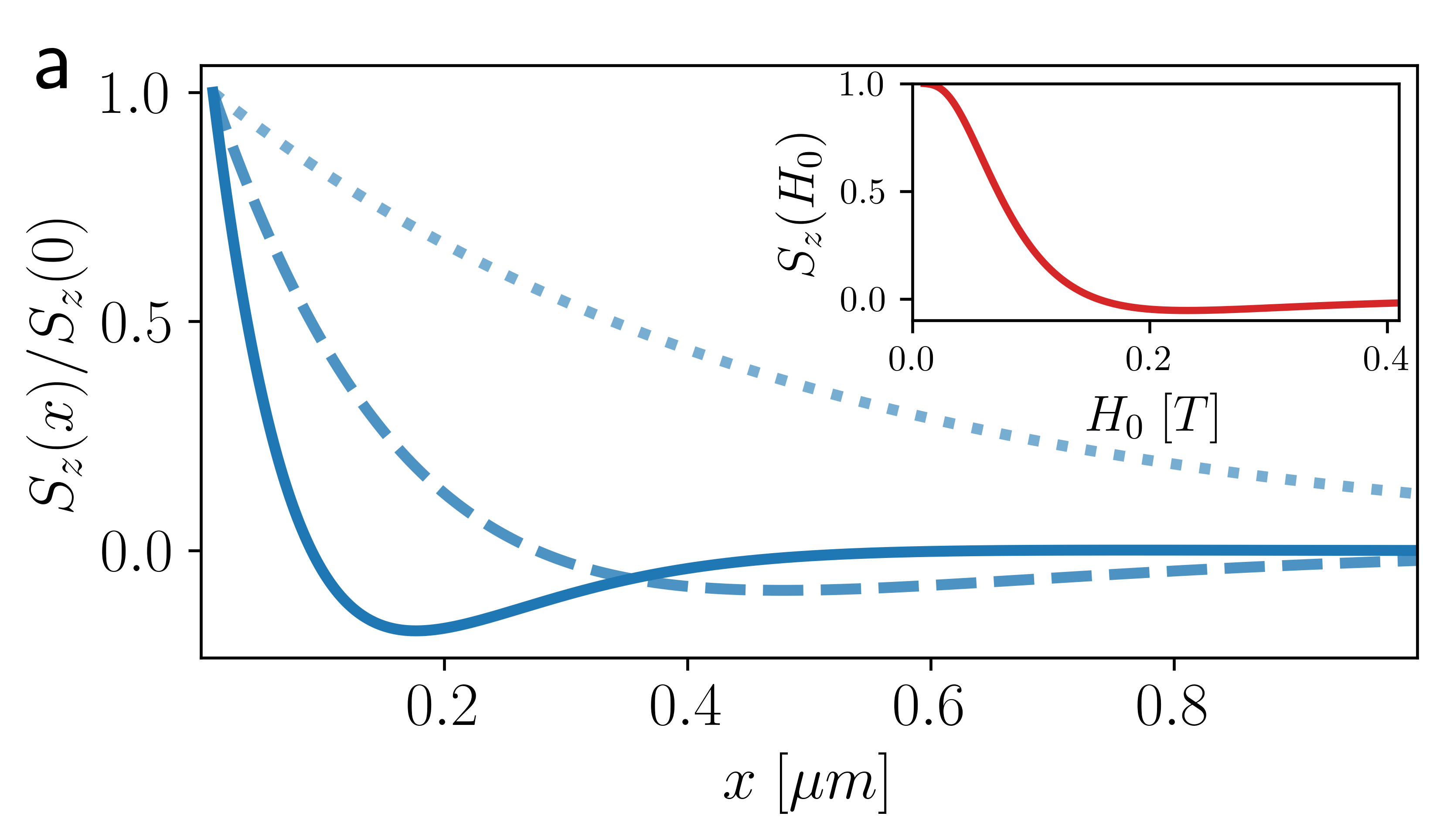}
	\includegraphics[width=.4\textwidth]{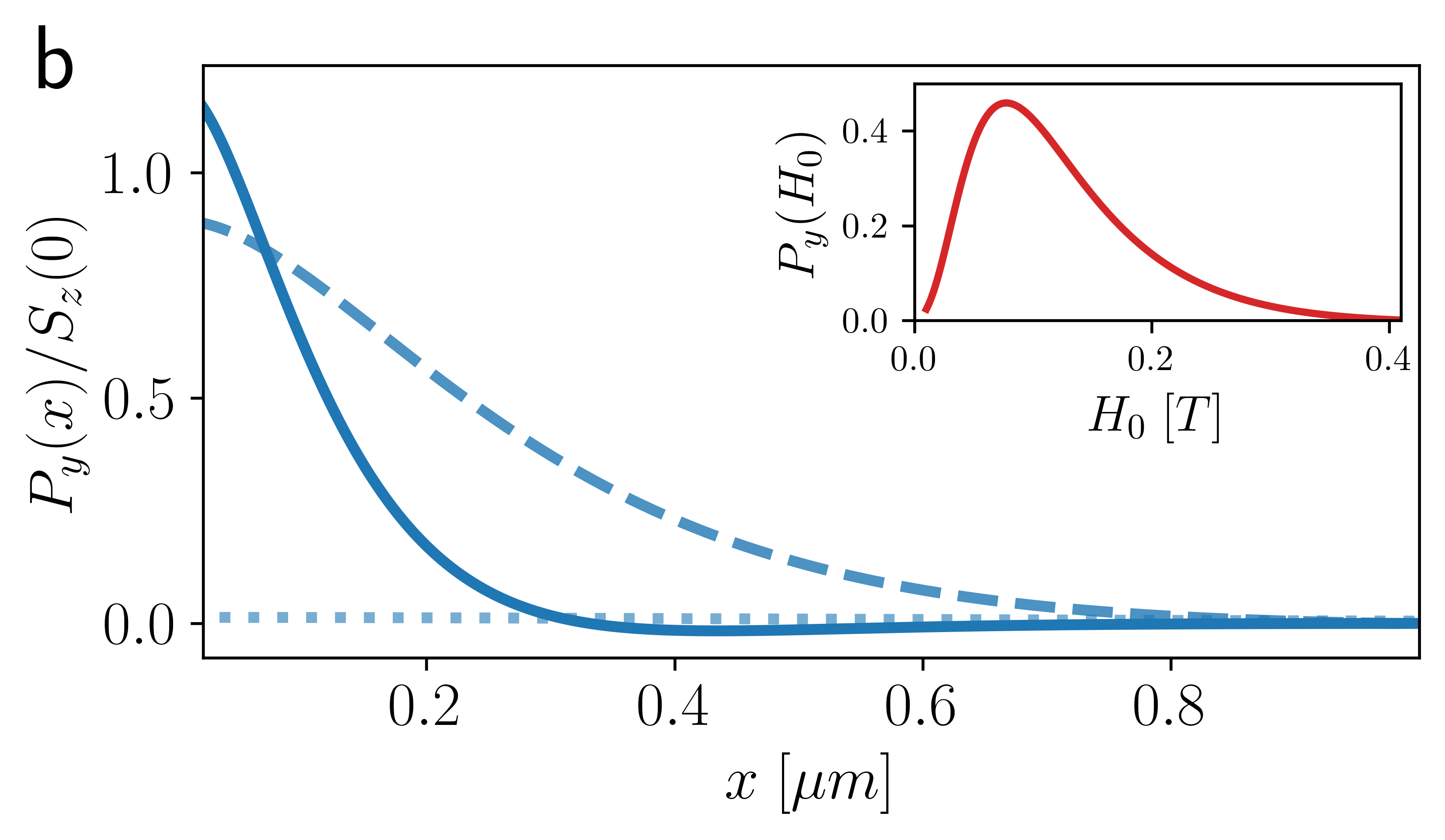}
	\includegraphics[width=.4\textwidth]{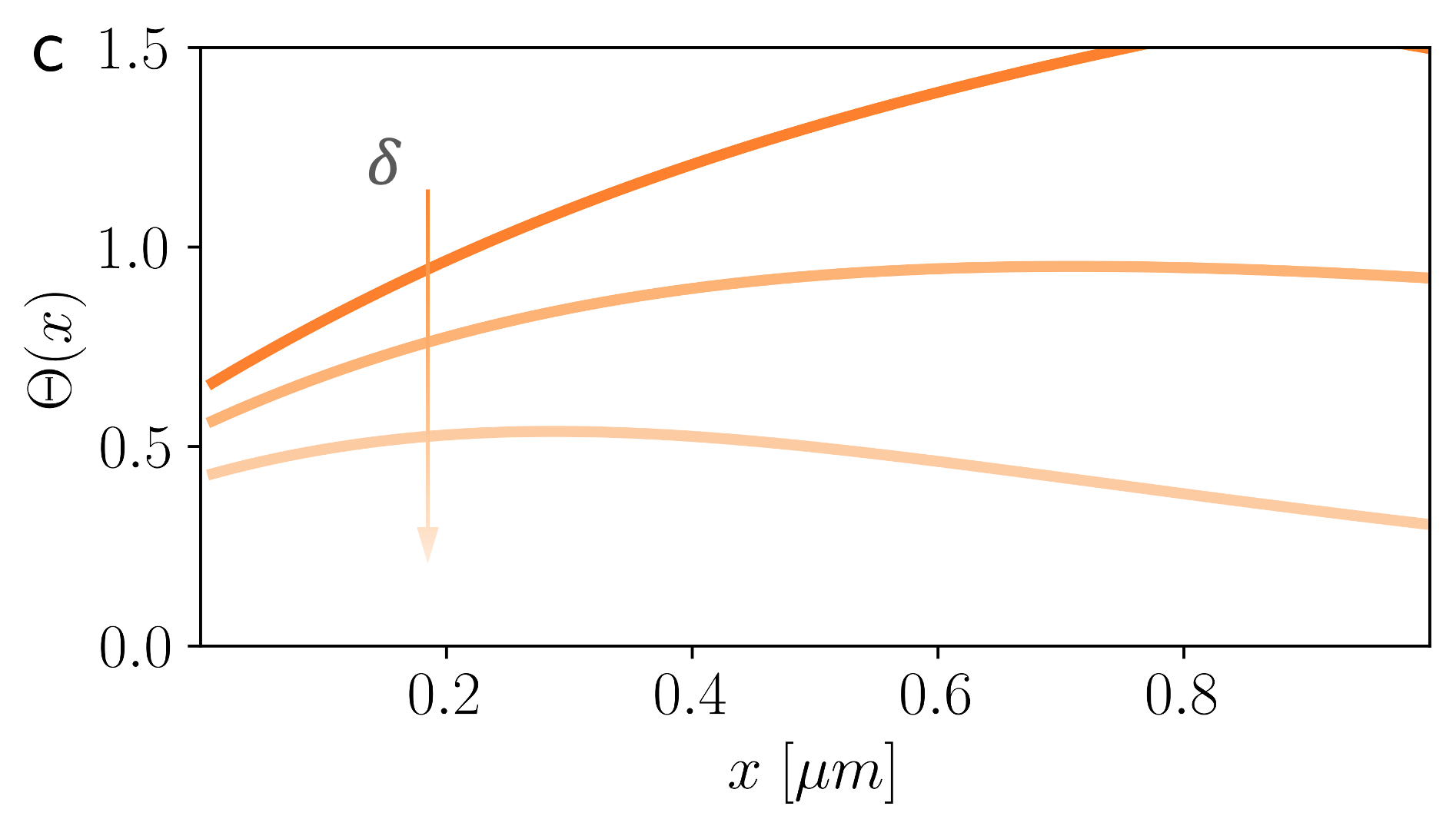} 
	\caption{The Hanle effect in the magnon spin accumulation $S_z(x)$ (a) and $P_y(x)$ (b). 
		Solid, dashed and dotted lines correspond to $B_0 = 0.8, 0.3,0.1$~T, respectively. 
		The insets show $S_z(B_0), P_y(B_0)$ at distance $x=0.5~\mu$m from an injector. 
		The parameters are taken for MnF$_2$,  $\tau_{g}=150$~ns, $D = 0.5 \times 10^{-5}$ m$^2$s$^{-1}$ and $T=20$~K. 
		(c) The spatial dependence of $\Theta(x) = {\rm arctan}(P_y/ S_z)$ with increasing $\delta$.} 
	\label{fig:sHanle}
\end{figure}

With a pseudospin texture, $\bm{\Omega}_{\bm k} \neq 0$, and its finite $\Gamma_x=\Gamma_y \neq \Gamma_z$, 
the eigenvalues of the static Eq.~(\ref{eq:DriftDiff}) are 
\begin{equation}
	\lambda_\pm^2(\varepsilon) = {L_\ast^{-2} \pm i {\tilde{F}_\varepsilon}/{D}},
\end{equation}
where $L_\ast^2 = D/\Gamma$ and $\tilde{F}_\varepsilon = \sqrt{F_\varepsilon^2 - \delta^2}$, 
with an average relaxation rate, $\Gamma = \tau_g^{-1} + (\Gamma_y + \Gamma_z)/2$, 
and an anisotropy factor, $\delta =  (\Gamma_y - \Gamma_z)/2$, respectively.

In Fig.~\ref{fig:sHanle}(a),(b) we show the static accumulation of $S_z(x)$ and $P_y(x)$, 
as a function of the distance from injector for different $B_0$~\cite{SM}. 
The parameters are taken for an easy-axis MnF$_2$ 
with $\bm{L} \parallel \hat{\bm{z}}$, $\bm{B}_0 \parallel \hat{\bm{x}}$, 
and $T$-range such that $\hat{\Gamma} \ll \tau_g$ for most magnons. 
$\bm{B}_0 \parallel \hat{\bm{x}} $ results in a fast decay of $S_z(x)$: 
$S_z(B_0)$ at a fixed $x$ is shown in the inset. 
A characteristic magnitude of $B_\ast$, at which the signal is fully suppressed, is typically below $1$~T [we have $B_{\ast} \sim 0.3$~T for Fig.~\ref{fig:sHanle}(a)].
In Fig.~\ref{fig:sHanle}(b) we reveal the accumulation of magnon linear polarization in $P_y(x)$, resulting from a strong conversion at $B_0\neq 0$.  $P_y(B_0)$ at a distance $x$, 
has a nonmonotonic character and reaches a maximum around $B_0 \sim 0.15$~T [inset in Fig.~\ref{fig:sHanle}(b)]. 
The magnon Hanle effect has been also discussed for easy-plane AFMs~\cite{kamra2020antiferromagnetic,wimmer2020observation,kleinherbers2024magnon,su2024controllable}, 
where the presence of both magnetic anisotropy field and $\bm{B}_0 \perp \bm{L}$ 
can shift the location of $S_z(B_0)$ maximum at finite $B_0$~\cite{kleinherbers2024magnon}.  

In addition, our theory reveals the role of the anisotropic DP relaxation for the magnon Hanle effect, which was previously ignored. 
We note that $\lambda_\pm$ is sensitive to the relaxation of 
the in-plane pseudospin components via $\Gamma_y$ and that both AMs and DM-equipped lattices 
provide us with an extreme anisotropy of $\Gamma_y \gg \Gamma_z$. 
First, this leads to the reduction in 
$\tilde{F}_\varepsilon$ at $\delta \neq 0$, 
which redefines a typical value of $B_\ast$ for the Hanle effect. 
Second, when $\tau_g^{-1} \ll \Gamma_y$, 
the DP relaxation is faster than Gilbert damping (expected to happen at low $T$)
resulting in the inefficient conversion of $S_z$ to $P_{x,y}$ due to its enhanced relaxation. 
This is illustrated in Fig.~\ref{fig:sHanle}(c), where we present the degree of ellipticity, 
$\Theta(x) = {\rm arctan}(P_y/ S_z)$,  as a function of distance from the injector for different $\delta$; 
there is a strong suppression of $\Theta$ as $\delta$ increases.

While in this work we focused on collinear AFMs, we envision that similar phenomena are present for any magnetically ordered systems carrying multiple spin excitation bands, such as in noncollinear and noncoplanar AFMs~\cite{okuma2017magnon,owerre2017topological,laurell2018magnon,doki2018spin}, including Kagome lattices and skyrmion crystals with finite magnon topology. 
Furthermore, we argue that the polarization-sensitive transport is attainable in systems there magnons hybridize with other quasiparticles, as phonons or excitons due to enhanced exciton-light-magnon coupling in van der Waals materials~\cite{bae2022exciton,wang2023magnetically}. 
The issue of magnon polarization transport and its conversion effect could stimulate advancements in detection techniques~\cite{kargar2021advances}, 
and serve to strengthen magnon-current functionalities that integrate nonvolatile memory and logic~\cite{liang2026ferroelectric}.

\textit{Acknowledgements}.
This work is supported by the U.S.
Department of Energy, Office of Science, Basic Energy
Sciences under Award No. DE-SC0004890. Computational
resources were provided by the UB Center for
Computational Research.

\textit{Data availability}. There are no publicly available research
data or software supporting this manuscript. Requests for further
information or data should be sent to the authors.

	\newpage
	
\bibliography{Ref}

\begin{thebibliography}{99}%
\makeatletter
\providecommand \@ifxundefined [1]{%
 \@ifx{#1\undefined}
}%
\providecommand \@ifnum [1]{%
 \ifnum #1\expandafter \@firstoftwo
 \else \expandafter \@secondoftwo
 \fi
}%
\providecommand \@ifx [1]{%
 \ifx #1\expandafter \@firstoftwo
 \else \expandafter \@secondoftwo
 \fi
}%
\providecommand \natexlab [1]{#1}%
\providecommand \enquote  [1]{``#1''}%
\providecommand \bibnamefont  [1]{#1}%
\providecommand \bibfnamefont [1]{#1}%
\providecommand \citenamefont [1]{#1}%
\providecommand \href@noop [0]{\@secondoftwo}%
\providecommand \href [0]{\begingroup \@sanitize@url \@href}%
\providecommand \@href[1]{\@@startlink{#1}\@@href}%
\providecommand \@@href[1]{\endgroup#1\@@endlink}%
\providecommand \@sanitize@url [0]{\catcode `\\12\catcode `\$12\catcode
  `\&12\catcode `\#12\catcode `\^12\catcode `\_12\catcode `\%12\relax}%
\providecommand \@@startlink[1]{}%
\providecommand \@@endlink[0]{}%
\providecommand \url  [0]{\begingroup\@sanitize@url \@url }%
\providecommand \@url [1]{\endgroup\@href {#1}{\urlprefix }}%
\providecommand \urlprefix  [0]{URL }%
\providecommand \Eprint [0]{\href }%
\providecommand \doibase [0]{https://doi.org/}%
\providecommand \selectlanguage [0]{\@gobble}%
\providecommand \bibinfo  [0]{\@secondoftwo}%
\providecommand \bibfield  [0]{\@secondoftwo}%
\providecommand \translation [1]{[#1]}%
\providecommand \BibitemOpen [0]{}%
\providecommand \bibitemStop [0]{}%
\providecommand \bibitemNoStop [0]{.\EOS\space}%
\providecommand \EOS [0]{\spacefactor3000\relax}%
\providecommand \BibitemShut  [1]{\csname bibitem#1\endcsname}%
\let\auto@bib@innerbib\@empty
\bibitem [{\citenamefont {Bloch}(1930)}]{bloch1930theorie}%
  \BibitemOpen
  \bibfield  {author} {\bibinfo {author} {\bibfnamefont {F.}~\bibnamefont
  {Bloch}},\ }\bibfield  {title} {\bibinfo {title} {Zur theorie des
  ferromagnetismus},\ }\href@noop {} {\bibfield  {journal} {\bibinfo  {journal}
  {Zeitschrift f{\"u}r Physik}\ }\textbf {\bibinfo {volume} {61}},\ \bibinfo
  {pages} {206} (\bibinfo {year} {1930})}\BibitemShut {NoStop}%
\bibitem [{\citenamefont {Kittel}\ and\ \citenamefont
  {Fong}(1963)}]{kittel1963quantum}%
  \BibitemOpen
  \bibfield  {author} {\bibinfo {author} {\bibfnamefont {C.}~\bibnamefont
  {Kittel}}\ and\ \bibinfo {author} {\bibfnamefont {C.-y.}\ \bibnamefont
  {Fong}},\ }\href@noop {} {\emph {\bibinfo {title} {Quantum theory of
  solids}}},\ Vol.~\bibinfo {volume} {5}\ (\bibinfo  {publisher} {Wiley New
  York},\ \bibinfo {year} {1963})\BibitemShut {NoStop}%
\bibitem [{\citenamefont {Kruglyak}\ \emph {et~al.}(2010)\citenamefont
  {Kruglyak}, \citenamefont {Demokritov},\ and\ \citenamefont
  {Grundler}}]{kruglyak2010magnonics}%
  \BibitemOpen
  \bibfield  {author} {\bibinfo {author} {\bibfnamefont {V.}~\bibnamefont
  {Kruglyak}}, \bibinfo {author} {\bibfnamefont {S.}~\bibnamefont
  {Demokritov}},\ and\ \bibinfo {author} {\bibfnamefont {D.}~\bibnamefont
  {Grundler}},\ }\bibfield  {title} {\bibinfo {title} {Magnonics},\ }\href
  {https://doi.org/10.1088/0022-3727/43/26/264001} {\bibfield  {journal}
  {\bibinfo  {journal} {J. Phys. D Appl. Phys.}\ }\textbf {\bibinfo {volume}
  {43}},\ \bibinfo {pages} {264001} (\bibinfo {year} {2010})}\BibitemShut
  {NoStop}%
\bibitem [{\citenamefont {Lenk}\ \emph {et~al.}(2011)\citenamefont {Lenk},
  \citenamefont {Ulrichs}, \citenamefont {Garbs},\ and\ \citenamefont
  {M{\"u}nzenberg}}]{lenk2011building}%
  \BibitemOpen
  \bibfield  {author} {\bibinfo {author} {\bibfnamefont {B.}~\bibnamefont
  {Lenk}}, \bibinfo {author} {\bibfnamefont {H.}~\bibnamefont {Ulrichs}},
  \bibinfo {author} {\bibfnamefont {F.}~\bibnamefont {Garbs}},\ and\ \bibinfo
  {author} {\bibfnamefont {M.}~\bibnamefont {M{\"u}nzenberg}},\ }\bibfield
  {title} {\bibinfo {title} {The building blocks of magnonics},\ }\href
  {https://doi.org/10.1016/j.physrep.2011.06.003} {\bibfield  {journal}
  {\bibinfo  {journal} {Phys. Rep.}\ }\textbf {\bibinfo {volume} {507}},\
  \bibinfo {pages} {107} (\bibinfo {year} {2011})}\BibitemShut {NoStop}%
\bibitem [{\citenamefont {Pirro}\ \emph {et~al.}(2021)\citenamefont {Pirro},
  \citenamefont {Vasyuchka}, \citenamefont {Serga},\ and\ \citenamefont
  {Hillebrands}}]{pirro2021advances}%
  \BibitemOpen
  \bibfield  {author} {\bibinfo {author} {\bibfnamefont {P.}~\bibnamefont
  {Pirro}}, \bibinfo {author} {\bibfnamefont {V.~I.}\ \bibnamefont
  {Vasyuchka}}, \bibinfo {author} {\bibfnamefont {A.~A.}\ \bibnamefont
  {Serga}},\ and\ \bibinfo {author} {\bibfnamefont {B.}~\bibnamefont
  {Hillebrands}},\ }\bibfield  {title} {\bibinfo {title} {Advances in coherent
  magnonics},\ }\href {https://doi.org/10.1038/s41578-021-00332-w} {\bibfield
  {journal} {\bibinfo  {journal} {Nat. Rev. Mater.}\ }\textbf {\bibinfo
  {volume} {6}},\ \bibinfo {pages} {1114} (\bibinfo {year} {2021})}\BibitemShut
  {NoStop}%
\bibitem [{\citenamefont {Flebus}\ \emph {et~al.}(2024)\citenamefont {Flebus},
  \citenamefont {Grundler}, \citenamefont {Rana}, \citenamefont {Otani},
  \citenamefont {Barsukov}, \citenamefont {Barman}, \citenamefont {Gubbiotti},
  \citenamefont {Landeros}, \citenamefont {Akerman}, \citenamefont {Ebels}
  \emph {et~al.}}]{flebus2024roadmap}%
  \BibitemOpen
  \bibfield  {author} {\bibinfo {author} {\bibfnamefont {B.}~\bibnamefont
  {Flebus}}, \bibinfo {author} {\bibfnamefont {D.}~\bibnamefont {Grundler}},
  \bibinfo {author} {\bibfnamefont {B.}~\bibnamefont {Rana}}, \bibinfo {author}
  {\bibfnamefont {Y.}~\bibnamefont {Otani}}, \bibinfo {author} {\bibfnamefont
  {I.}~\bibnamefont {Barsukov}}, \bibinfo {author} {\bibfnamefont
  {A.}~\bibnamefont {Barman}}, \bibinfo {author} {\bibfnamefont
  {G.}~\bibnamefont {Gubbiotti}}, \bibinfo {author} {\bibfnamefont
  {P.}~\bibnamefont {Landeros}}, \bibinfo {author} {\bibfnamefont
  {J.}~\bibnamefont {Akerman}}, \bibinfo {author} {\bibfnamefont
  {U.}~\bibnamefont {Ebels}}, \emph {et~al.},\ }\bibfield  {title} {\bibinfo
  {title} {The 2024 magnonics roadmap},\ }\href
  {https://doi.org/10.1088/1361-648X/ad399c} {\bibfield  {journal} {\bibinfo
  {journal} {J. Phys. Condens. Matter}\ }\textbf {\bibinfo {volume} {36}},\
  \bibinfo {pages} {363501} (\bibinfo {year} {2024})}\BibitemShut {NoStop}%
\bibitem [{\citenamefont {{\v{Z}}uti{\'c}}\ \emph {et~al.}(2004)\citenamefont
  {{\v{Z}}uti{\'c}}, \citenamefont {Fabian},\ and\ \citenamefont
  {Sarma}}]{vzutic2004spintronics}%
  \BibitemOpen
  \bibfield  {author} {\bibinfo {author} {\bibfnamefont {I.}~\bibnamefont
  {{\v{Z}}uti{\'c}}}, \bibinfo {author} {\bibfnamefont {J.}~\bibnamefont
  {Fabian}},\ and\ \bibinfo {author} {\bibfnamefont {S.~D.}\ \bibnamefont
  {Sarma}},\ }\bibfield  {title} {\bibinfo {title} {Spintronics: Fundamentals
  and applications},\ }\href {https://doi.org/10.1103/RevModPhys.76.323}
  {\bibfield  {journal} {\bibinfo  {journal} {Rev. Mod. Phys.}\ }\textbf
  {\bibinfo {volume} {76}},\ \bibinfo {pages} {323} (\bibinfo {year}
  {2004})}\BibitemShut {NoStop}%
\bibitem [{\citenamefont {Tsymbal}\ and\ \citenamefont
  {\v{Z}uti\'c}(2019)}]{Tsymbal:2019}%
  \BibitemOpen
  \bibinfo {editor} {\bibfnamefont {E.~Y.}\ \bibnamefont {Tsymbal}}\ and\
  \bibinfo {editor} {\bibfnamefont {I.}~\bibnamefont {\v{Z}uti\'c}},\ eds.,\
  \href@noop {} {\emph {\bibinfo {title} {Spintronics Handbook Spin Transport
  and Magnetism, 2nd Edition}}}\ (\bibinfo  {publisher} {CRC Press, Taylor \&
  Francis, Boca Raton, FL},\ \bibinfo {year} {2019})\BibitemShut {NoStop}%
\bibitem [{\citenamefont {Chumak}\ \emph {et~al.}(2014)\citenamefont {Chumak},
  \citenamefont {Serga},\ and\ \citenamefont {Hillebrands}}]{chumak2014magnon}%
  \BibitemOpen
  \bibfield  {author} {\bibinfo {author} {\bibfnamefont {A.~V.}\ \bibnamefont
  {Chumak}}, \bibinfo {author} {\bibfnamefont {A.~A.}\ \bibnamefont {Serga}},\
  and\ \bibinfo {author} {\bibfnamefont {B.}~\bibnamefont {Hillebrands}},\
  }\bibfield  {title} {\bibinfo {title} {Magnon transistor for all-magnon data
  processing},\ }\href {https://doi.org/10.1038/ncomms5700} {\bibfield
  {journal} {\bibinfo  {journal} {Natur. Commun.}\ }\textbf {\bibinfo {volume}
  {5}},\ \bibinfo {pages} {4700} (\bibinfo {year} {2014})}\BibitemShut
  {NoStop}%
\bibitem [{\citenamefont {Chumak}\ \emph {et~al.}(2022)\citenamefont {Chumak},
  \citenamefont {Kabos}, \citenamefont {Wu}, \citenamefont {Abert},
  \citenamefont {Adelmann}, \citenamefont {Adeyeye}, \citenamefont
  {{\AA}kerman}, \citenamefont {Aliev}, \citenamefont {Anane}, \citenamefont
  {Awad} \emph {et~al.}}]{chumak2022advances}%
  \BibitemOpen
  \bibfield  {author} {\bibinfo {author} {\bibfnamefont {A.~V.}\ \bibnamefont
  {Chumak}}, \bibinfo {author} {\bibfnamefont {P.}~\bibnamefont {Kabos}},
  \bibinfo {author} {\bibfnamefont {M.}~\bibnamefont {Wu}}, \bibinfo {author}
  {\bibfnamefont {C.}~\bibnamefont {Abert}}, \bibinfo {author} {\bibfnamefont
  {C.}~\bibnamefont {Adelmann}}, \bibinfo {author} {\bibfnamefont {A.~O.}\
  \bibnamefont {Adeyeye}}, \bibinfo {author} {\bibfnamefont {J.}~\bibnamefont
  {{\AA}kerman}}, \bibinfo {author} {\bibfnamefont {F.~G.}\ \bibnamefont
  {Aliev}}, \bibinfo {author} {\bibfnamefont {A.}~\bibnamefont {Anane}},
  \bibinfo {author} {\bibfnamefont {A.}~\bibnamefont {Awad}}, \emph {et~al.},\
  }\bibfield  {title} {\bibinfo {title} {Advances in magnetics roadmap on
  spin-wave computing},\ }\href {https://doi.org/10.1109/TMAG.2022.3149664}
  {\bibfield  {journal} {\bibinfo  {journal} {IEEE Trans. Magn.}\ }\textbf
  {\bibinfo {volume} {58}},\ \bibinfo {pages} {1} (\bibinfo {year}
  {2022})}\BibitemShut {NoStop}%
\bibitem [{\citenamefont {Lachance-Quirion}\ \emph {et~al.}(2019)\citenamefont
  {Lachance-Quirion}, \citenamefont {Tabuchi}, \citenamefont {Gloppe},
  \citenamefont {Usami},\ and\ \citenamefont {Nakamura}}]{lachance2019hybrid}%
  \BibitemOpen
  \bibfield  {author} {\bibinfo {author} {\bibfnamefont {D.}~\bibnamefont
  {Lachance-Quirion}}, \bibinfo {author} {\bibfnamefont {Y.}~\bibnamefont
  {Tabuchi}}, \bibinfo {author} {\bibfnamefont {A.}~\bibnamefont {Gloppe}},
  \bibinfo {author} {\bibfnamefont {K.}~\bibnamefont {Usami}},\ and\ \bibinfo
  {author} {\bibfnamefont {Y.}~\bibnamefont {Nakamura}},\ }\bibfield  {title}
  {\bibinfo {title} {Hybrid quantum systems based on magnonics},\ }\href
  {https://doi.org/10.7567/1882-0786/ab248d} {\bibfield  {journal} {\bibinfo
  {journal} {Appl. Phys. Expr.}\ }\textbf {\bibinfo {volume} {12}},\ \bibinfo
  {pages} {070101} (\bibinfo {year} {2019})}\BibitemShut {NoStop}%
\bibitem [{\citenamefont {Li}\ \emph {et~al.}(2020)\citenamefont {Li},
  \citenamefont {Zhang}, \citenamefont {Tyberkevych}, \citenamefont {Kwok},
  \citenamefont {Hoffmann},\ and\ \citenamefont {Novosad}}]{li2020hybrid}%
  \BibitemOpen
  \bibfield  {author} {\bibinfo {author} {\bibfnamefont {Y.}~\bibnamefont
  {Li}}, \bibinfo {author} {\bibfnamefont {W.}~\bibnamefont {Zhang}}, \bibinfo
  {author} {\bibfnamefont {V.}~\bibnamefont {Tyberkevych}}, \bibinfo {author}
  {\bibfnamefont {W.-K.}\ \bibnamefont {Kwok}}, \bibinfo {author}
  {\bibfnamefont {A.}~\bibnamefont {Hoffmann}},\ and\ \bibinfo {author}
  {\bibfnamefont {V.}~\bibnamefont {Novosad}},\ }\bibfield  {title} {\bibinfo
  {title} {Hybrid magnonics: {Physics}, circuits, and applications for coherent
  information processing},\ }\bibfield  {journal} {\bibinfo  {journal} {J.
  Appl. Phys.}\ }\textbf {\bibinfo {volume} {128}},\ \href
  {https://doi.org/10.1063/5.0020277} {10.1063/5.0020277} (\bibinfo {year}
  {2020})\BibitemShut {NoStop}%
\bibitem [{\citenamefont {Serha}\ \emph {et~al.}(2026)\citenamefont {Serha},
  \citenamefont {Dubs},\ and\ \citenamefont {Chumak}}]{serha2026magnetic}%
  \BibitemOpen
  \bibfield  {author} {\bibinfo {author} {\bibfnamefont {R.~O.}\ \bibnamefont
  {Serha}}, \bibinfo {author} {\bibfnamefont {C.}~\bibnamefont {Dubs}},\ and\
  \bibinfo {author} {\bibfnamefont {A.~V.}\ \bibnamefont {Chumak}},\ }\bibfield
   {title} {\bibinfo {title} {Magnetic materials for quantum magnonics},\
  }\bibfield  {journal} {\bibinfo  {journal} {APL Materials}\ }\textbf
  {\bibinfo {volume} {14}},\ \href {https://doi.org/10.1063/5.0306423}
  {10.1063/5.0306423} (\bibinfo {year} {2026})\BibitemShut {NoStop}%
\bibitem [{\citenamefont {Baltz}\ \emph {et~al.}(2018)\citenamefont {Baltz},
  \citenamefont {Manchon}, \citenamefont {Tsoi}, \citenamefont {Moriyama},
  \citenamefont {Ono},\ and\ \citenamefont
  {Tserkovnyak}}]{baltz2018antiferromagnetic}%
  \BibitemOpen
  \bibfield  {author} {\bibinfo {author} {\bibfnamefont {V.}~\bibnamefont
  {Baltz}}, \bibinfo {author} {\bibfnamefont {A.}~\bibnamefont {Manchon}},
  \bibinfo {author} {\bibfnamefont {M.}~\bibnamefont {Tsoi}}, \bibinfo {author}
  {\bibfnamefont {T.}~\bibnamefont {Moriyama}}, \bibinfo {author}
  {\bibfnamefont {T.}~\bibnamefont {Ono}},\ and\ \bibinfo {author}
  {\bibfnamefont {Y.}~\bibnamefont {Tserkovnyak}},\ }\bibfield  {title}
  {\bibinfo {title} {Antiferromagnetic spintronics},\ }\href
  {https://doi.org/10.1103/RevModPhys.90.015005} {\bibfield  {journal}
  {\bibinfo  {journal} {Rev. Mod. Phys.}\ }\textbf {\bibinfo {volume} {90}},\
  \bibinfo {pages} {015005} (\bibinfo {year} {2018})}\BibitemShut {NoStop}%
\bibitem [{\citenamefont {Hortensius}\ \emph {et~al.}(2021)\citenamefont
  {Hortensius}, \citenamefont {Afanasiev}, \citenamefont {Matthiesen},
  \citenamefont {Leenders}, \citenamefont {Citro}, \citenamefont {Kimel},
  \citenamefont {Mikhaylovskiy}, \citenamefont {Ivanov},\ and\ \citenamefont
  {Caviglia}}]{hortensius2021coherent}%
  \BibitemOpen
  \bibfield  {author} {\bibinfo {author} {\bibfnamefont {J.}~\bibnamefont
  {Hortensius}}, \bibinfo {author} {\bibfnamefont {D.}~\bibnamefont
  {Afanasiev}}, \bibinfo {author} {\bibfnamefont {M.}~\bibnamefont
  {Matthiesen}}, \bibinfo {author} {\bibfnamefont {R.}~\bibnamefont
  {Leenders}}, \bibinfo {author} {\bibfnamefont {R.}~\bibnamefont {Citro}},
  \bibinfo {author} {\bibfnamefont {A.}~\bibnamefont {Kimel}}, \bibinfo
  {author} {\bibfnamefont {R.}~\bibnamefont {Mikhaylovskiy}}, \bibinfo {author}
  {\bibfnamefont {B.}~\bibnamefont {Ivanov}},\ and\ \bibinfo {author}
  {\bibfnamefont {A.}~\bibnamefont {Caviglia}},\ }\bibfield  {title} {\bibinfo
  {title} {Coherent spin-wave transport in an antiferromagnet},\ }\href
  {https://doi.org/10.1038/s41567-021-01290-4} {\bibfield  {journal} {\bibinfo
  {journal} {Nat. Phys.}\ }\textbf {\bibinfo {volume} {17}},\ \bibinfo {pages}
  {1001} (\bibinfo {year} {2021})}\BibitemShut {NoStop}%
\bibitem [{\citenamefont {Han}\ \emph {et~al.}(2023)\citenamefont {Han},
  \citenamefont {Cheng}, \citenamefont {Liu}, \citenamefont {Ohno},\ and\
  \citenamefont {Fukami}}]{han2023coherent}%
  \BibitemOpen
  \bibfield  {author} {\bibinfo {author} {\bibfnamefont {J.}~\bibnamefont
  {Han}}, \bibinfo {author} {\bibfnamefont {R.}~\bibnamefont {Cheng}}, \bibinfo
  {author} {\bibfnamefont {L.}~\bibnamefont {Liu}}, \bibinfo {author}
  {\bibfnamefont {H.}~\bibnamefont {Ohno}},\ and\ \bibinfo {author}
  {\bibfnamefont {S.}~\bibnamefont {Fukami}},\ }\bibfield  {title} {\bibinfo
  {title} {Coherent antiferromagnetic spintronics},\ }\href
  {https://doi.org/10.1038/s41563-023-01492-6} {\bibfield  {journal} {\bibinfo
  {journal} {Nat. Mater.}\ }\textbf {\bibinfo {volume} {22}},\ \bibinfo {pages}
  {684} (\bibinfo {year} {2023})}\BibitemShut {NoStop}%
\bibitem [{\citenamefont {Rimmler}\ \emph {et~al.}(2025)\citenamefont
  {Rimmler}, \citenamefont {Pal},\ and\ \citenamefont
  {Parkin}}]{rimmler2025non}%
  \BibitemOpen
  \bibfield  {author} {\bibinfo {author} {\bibfnamefont {B.~H.}\ \bibnamefont
  {Rimmler}}, \bibinfo {author} {\bibfnamefont {B.}~\bibnamefont {Pal}},\ and\
  \bibinfo {author} {\bibfnamefont {S.~S.}\ \bibnamefont {Parkin}},\ }\bibfield
   {title} {\bibinfo {title} {Non-collinear antiferromagnetic spintronics},\
  }\href {https://doi.org/10.1038/s41578-024-00706-w} {\bibfield  {journal}
  {\bibinfo  {journal} {Natur. Rev. Mater.}\ }\textbf {\bibinfo {volume}
  {10}},\ \bibinfo {pages} {109} (\bibinfo {year} {2025})}\BibitemShut
  {NoStop}%
\bibitem [{\citenamefont {Shiota}\ \emph {et~al.}(2024)\citenamefont {Shiota},
  \citenamefont {Taniguchi}, \citenamefont {Hayashi}, \citenamefont {Narita},
  \citenamefont {Karube}, \citenamefont {Hisatomi}, \citenamefont {Moriyama},\
  and\ \citenamefont {Ono}}]{shiota2024handedness}%
  \BibitemOpen
  \bibfield  {author} {\bibinfo {author} {\bibfnamefont {Y.}~\bibnamefont
  {Shiota}}, \bibinfo {author} {\bibfnamefont {T.}~\bibnamefont {Taniguchi}},
  \bibinfo {author} {\bibfnamefont {D.}~\bibnamefont {Hayashi}}, \bibinfo
  {author} {\bibfnamefont {H.}~\bibnamefont {Narita}}, \bibinfo {author}
  {\bibfnamefont {S.}~\bibnamefont {Karube}}, \bibinfo {author} {\bibfnamefont
  {R.}~\bibnamefont {Hisatomi}}, \bibinfo {author} {\bibfnamefont
  {T.}~\bibnamefont {Moriyama}},\ and\ \bibinfo {author} {\bibfnamefont
  {T.}~\bibnamefont {Ono}},\ }\bibfield  {title} {\bibinfo {title} {Handedness
  manipulation of propagating antiferromagnetic magnons},\ }\href
  {https://doi.org/10.1038/s41467-024-54125-0} {\bibfield  {journal} {\bibinfo
  {journal} {Nat. Comm.}\ }\textbf {\bibinfo {volume} {15}},\ \bibinfo {pages}
  {9750} (\bibinfo {year} {2024})}\BibitemShut {NoStop}%
\bibitem [{\citenamefont {Sun}\ \emph {et~al.}(2024)\citenamefont {Sun},
  \citenamefont {Meng}, \citenamefont {Lee}, \citenamefont {Soll},
  \citenamefont {Zhang}, \citenamefont {Ramesh}, \citenamefont {Yao},
  \citenamefont {Sofer},\ and\ \citenamefont {Orenstein}}]{sun2024dipolar}%
  \BibitemOpen
  \bibfield  {author} {\bibinfo {author} {\bibfnamefont {Y.}~\bibnamefont
  {Sun}}, \bibinfo {author} {\bibfnamefont {F.}~\bibnamefont {Meng}}, \bibinfo
  {author} {\bibfnamefont {C.}~\bibnamefont {Lee}}, \bibinfo {author}
  {\bibfnamefont {A.}~\bibnamefont {Soll}}, \bibinfo {author} {\bibfnamefont
  {H.}~\bibnamefont {Zhang}}, \bibinfo {author} {\bibfnamefont
  {R.}~\bibnamefont {Ramesh}}, \bibinfo {author} {\bibfnamefont
  {J.}~\bibnamefont {Yao}}, \bibinfo {author} {\bibfnamefont {Z.}~\bibnamefont
  {Sofer}},\ and\ \bibinfo {author} {\bibfnamefont {J.}~\bibnamefont
  {Orenstein}},\ }\bibfield  {title} {\bibinfo {title} {Dipolar spin wave
  packet transport in a van der {Waals} antiferromagnet},\ }\href
  {https://doi.org/10.1038/s41567-024-02387-2} {\bibfield  {journal} {\bibinfo
  {journal} {Nat. Phys.}\ }\textbf {\bibinfo {volume} {20}},\ \bibinfo {pages}
  {794} (\bibinfo {year} {2024})}\BibitemShut {NoStop}%
\bibitem [{\citenamefont {Leenders}\ \emph {et~al.}(2024)\citenamefont
  {Leenders}, \citenamefont {Afanasiev}, \citenamefont {Kimel},\ and\
  \citenamefont {Mikhaylovskiy}}]{leenders2024canted}%
  \BibitemOpen
  \bibfield  {author} {\bibinfo {author} {\bibfnamefont {R.~A.}\ \bibnamefont
  {Leenders}}, \bibinfo {author} {\bibfnamefont {D.}~\bibnamefont {Afanasiev}},
  \bibinfo {author} {\bibfnamefont {A.~V.}\ \bibnamefont {Kimel}},\ and\
  \bibinfo {author} {\bibfnamefont {R.~V.}\ \bibnamefont {Mikhaylovskiy}},\
  }\bibfield  {title} {\bibinfo {title} {Canted spin order as a platform for
  ultrafast conversion of magnons},\ }\href
  {https://doi.org/10.1038/s41586-024-07448-3} {\bibfield  {journal} {\bibinfo
  {journal} {Nature}\ }\textbf {\bibinfo {volume} {630}},\ \bibinfo {pages}
  {335} (\bibinfo {year} {2024})}\BibitemShut {NoStop}%
\bibitem [{\citenamefont {Okuma}(2017)}]{okuma2017magnon}%
  \BibitemOpen
  \bibfield  {author} {\bibinfo {author} {\bibfnamefont {N.}~\bibnamefont
  {Okuma}},\ }\bibfield  {title} {\bibinfo {title} {{Magnon Spin-Momentum
  Locking: Various Spin Vortices and Dirac magnons in Noncollinear
  Antiferromagnets}},\ }\href {https://doi.org/10.1103/PhysRevLett.119.107205}
  {\bibfield  {journal} {\bibinfo  {journal} {Phys. Rev. Lett.}\ }\textbf
  {\bibinfo {volume} {119}},\ \bibinfo {pages} {107205} (\bibinfo {year}
  {2017})}\BibitemShut {NoStop}%
\bibitem [{\citenamefont {Kawano}\ \emph {et~al.}(2019)\citenamefont {Kawano},
  \citenamefont {Onose},\ and\ \citenamefont {Hotta}}]{kawano2019designing}%
  \BibitemOpen
  \bibfield  {author} {\bibinfo {author} {\bibfnamefont {M.}~\bibnamefont
  {Kawano}}, \bibinfo {author} {\bibfnamefont {Y.}~\bibnamefont {Onose}},\ and\
  \bibinfo {author} {\bibfnamefont {C.}~\bibnamefont {Hotta}},\ }\bibfield
  {title} {\bibinfo {title} {Designing {Rashba}-{Dresselhaus} effect in
  magnetic insulators},\ }\href {https://doi.org/10.1038/s42005-019-0128-6}
  {\bibfield  {journal} {\bibinfo  {journal} {Commun. Phys.}\ }\textbf
  {\bibinfo {volume} {2}},\ \bibinfo {pages} {27} (\bibinfo {year}
  {2019})}\BibitemShut {NoStop}%
\bibitem [{\citenamefont {Kawano}\ and\ \citenamefont
  {Hotta}(2019{\natexlab{a}})}]{kawano2019discovering}%
  \BibitemOpen
  \bibfield  {author} {\bibinfo {author} {\bibfnamefont {M.}~\bibnamefont
  {Kawano}}\ and\ \bibinfo {author} {\bibfnamefont {C.}~\bibnamefont {Hotta}},\
  }\bibfield  {title} {\bibinfo {title} {Discovering momentum-dependent magnon
  spin texture in insulating antiferromagnets: {Role} of the {Kitaev}
  interaction},\ }\href {https://doi.org/10.1103/PhysRevB.100.174402}
  {\bibfield  {journal} {\bibinfo  {journal} {Phys. Rev. B}\ }\textbf {\bibinfo
  {volume} {100}},\ \bibinfo {pages} {174402} (\bibinfo {year}
  {2019}{\natexlab{a}})}\BibitemShut {NoStop}%
\bibitem [{\citenamefont {Shen}(2020)}]{shen2020magnon}%
  \BibitemOpen
  \bibfield  {author} {\bibinfo {author} {\bibfnamefont {K.}~\bibnamefont
  {Shen}},\ }\bibfield  {title} {\bibinfo {title} {Magnon spin relaxation and
  spin {Hall} effect due to the dipolar interaction in antiferromagnetic
  insulators},\ }\href {https://doi.org/10.1103/PhysRevLett.124.077201}
  {\bibfield  {journal} {\bibinfo  {journal} {Phys. Rev. Lett.}\ }\textbf
  {\bibinfo {volume} {124}},\ \bibinfo {pages} {077201} (\bibinfo {year}
  {2020})}\BibitemShut {NoStop}%
\bibitem [{\citenamefont {Hu}\ \emph {et~al.}(2026)\citenamefont {Hu},
  \citenamefont {Wan}, \citenamefont {Savrasov},\ and\ \citenamefont
  {Liu}}]{hu2026anisotropic}%
  \BibitemOpen
  \bibfield  {author} {\bibinfo {author} {\bibfnamefont {Y.}~\bibnamefont
  {Hu}}, \bibinfo {author} {\bibfnamefont {X.}~\bibnamefont {Wan}}, \bibinfo
  {author} {\bibfnamefont {S.~Y.}\ \bibnamefont {Savrasov}},\ and\ \bibinfo
  {author} {\bibfnamefont {Y.}~\bibnamefont {Liu}},\ }\bibfield  {title}
  {\bibinfo {title} {Anisotropic symmetric {$\hat{\Gamma}$} -interaction
  induced magnon {Dyakonov-Perel and Elliott-Yafet} spin relaxation in
  two-dimensional antiferromagnets},\ }\href
  {https://doi.org/10.1103/1wgf-d8fp} {\bibfield  {journal} {\bibinfo
  {journal} {Phys. Rev. B}\ }\textbf {\bibinfo {volume} {113}},\ \bibinfo
  {pages} {024437} (\bibinfo {year} {2026})}\BibitemShut {NoStop}%
\bibitem [{\citenamefont {Gao}\ \emph {et~al.}(2026)\citenamefont {Gao},
  \citenamefont {Cong}, \citenamefont {Liang}, \citenamefont {Yang},
  \citenamefont {Liu}, \citenamefont {Chen}, \citenamefont {Wu}, \citenamefont
  {Shen},\ and\ \citenamefont {Li}}]{gao2026observation}%
  \BibitemOpen
  \bibfield  {author} {\bibinfo {author} {\bibfnamefont {Q.}~\bibnamefont
  {Gao}}, \bibinfo {author} {\bibfnamefont {A.}~\bibnamefont {Cong}}, \bibinfo
  {author} {\bibfnamefont {B.}~\bibnamefont {Liang}}, \bibinfo {author}
  {\bibfnamefont {M.}~\bibnamefont {Yang}}, \bibinfo {author} {\bibfnamefont
  {J.}~\bibnamefont {Liu}}, \bibinfo {author} {\bibfnamefont {L.}~\bibnamefont
  {Chen}}, \bibinfo {author} {\bibfnamefont {S.}~\bibnamefont {Wu}}, \bibinfo
  {author} {\bibfnamefont {K.}~\bibnamefont {Shen}},\ and\ \bibinfo {author}
  {\bibfnamefont {J.}~\bibnamefont {Li}},\ }\bibfield  {title} {\bibinfo
  {title} {Observation of {D'yakonov-Perel'}-type magnon spin relaxation in
  uniaxial antiferromagnetic insulators},\ }\bibfield  {journal} {\bibinfo
  {journal} {Nat. Commun.}\ }\textbf {\bibinfo {volume} {17}},\ \href
  {https://doi.org/10.1038/s41467-026-71298-y} {10.1038/s41467-026-71298-y}
  (\bibinfo {year} {2026})\BibitemShut {NoStop}%
\bibitem [{\citenamefont {Daniels}\ \emph {et~al.}(2018)\citenamefont
  {Daniels}, \citenamefont {Cheng}, \citenamefont {Yu}, \citenamefont {Xiao},\
  and\ \citenamefont {Xiao}}]{daniels2018nonabelian}%
  \BibitemOpen
  \bibfield  {author} {\bibinfo {author} {\bibfnamefont {M.~W.}\ \bibnamefont
  {Daniels}}, \bibinfo {author} {\bibfnamefont {R.}~\bibnamefont {Cheng}},
  \bibinfo {author} {\bibfnamefont {W.}~\bibnamefont {Yu}}, \bibinfo {author}
  {\bibfnamefont {J.}~\bibnamefont {Xiao}},\ and\ \bibinfo {author}
  {\bibfnamefont {D.}~\bibnamefont {Xiao}},\ }\bibfield  {title} {\bibinfo
  {title} {Nonabelian magnonics in antiferromagnets},\ }\href
  {https://doi.org/10.1103/PhysRevB.98.134450} {\bibfield  {journal} {\bibinfo
  {journal} {Phys. Rev. B}\ }\textbf {\bibinfo {volume} {98}},\ \bibinfo
  {pages} {134450} (\bibinfo {year} {2018})}\BibitemShut {NoStop}%
\bibitem [{\citenamefont {Wimmer}\ \emph {et~al.}(2020)\citenamefont {Wimmer},
  \citenamefont {Kamra}, \citenamefont {G{\"u}ckelhorn}, \citenamefont {Opel},
  \citenamefont {Gepr{\"a}gs}, \citenamefont {Gross}, \citenamefont {Huebl},\
  and\ \citenamefont {Althammer}}]{wimmer2020observation}%
  \BibitemOpen
  \bibfield  {author} {\bibinfo {author} {\bibfnamefont {T.}~\bibnamefont
  {Wimmer}}, \bibinfo {author} {\bibfnamefont {A.}~\bibnamefont {Kamra}},
  \bibinfo {author} {\bibfnamefont {J.}~\bibnamefont {G{\"u}ckelhorn}},
  \bibinfo {author} {\bibfnamefont {M.}~\bibnamefont {Opel}}, \bibinfo {author}
  {\bibfnamefont {S.}~\bibnamefont {Gepr{\"a}gs}}, \bibinfo {author}
  {\bibfnamefont {R.}~\bibnamefont {Gross}}, \bibinfo {author} {\bibfnamefont
  {H.}~\bibnamefont {Huebl}},\ and\ \bibinfo {author} {\bibfnamefont
  {M.}~\bibnamefont {Althammer}},\ }\bibfield  {title} {\bibinfo {title}
  {Observation of antiferromagnetic magnon pseudospin dynamics and the {Hanle}
  effect},\ }\href {https://doi.org/10.1103/PhysRevLett.125.247204} {\bibfield
  {journal} {\bibinfo  {journal} {Phys. Rev. Lett.}\ }\textbf {\bibinfo
  {volume} {125}},\ \bibinfo {pages} {247204} (\bibinfo {year}
  {2020})}\BibitemShut {NoStop}%
\bibitem [{\citenamefont {Kamra}\ \emph {et~al.}(2020)\citenamefont {Kamra},
  \citenamefont {Wimmer}, \citenamefont {Huebl},\ and\ \citenamefont
  {Althammer}}]{kamra2020antiferromagnetic}%
  \BibitemOpen
  \bibfield  {author} {\bibinfo {author} {\bibfnamefont {A.}~\bibnamefont
  {Kamra}}, \bibinfo {author} {\bibfnamefont {T.}~\bibnamefont {Wimmer}},
  \bibinfo {author} {\bibfnamefont {H.}~\bibnamefont {Huebl}},\ and\ \bibinfo
  {author} {\bibfnamefont {M.}~\bibnamefont {Althammer}},\ }\bibfield  {title}
  {\bibinfo {title} {Antiferromagnetic magnon pseudospin: {Dynamics} and
  diffusive transport},\ }\href {https://doi.org/10.1103/PhysRevB.102.174445}
  {\bibfield  {journal} {\bibinfo  {journal} {Phys. Rev. B}\ }\textbf {\bibinfo
  {volume} {102}},\ \bibinfo {pages} {174445} (\bibinfo {year}
  {2020})}\BibitemShut {NoStop}%
\bibitem [{\citenamefont {G{\"u}ckelhorn}\ \emph {et~al.}(2023)\citenamefont
  {G{\"u}ckelhorn}, \citenamefont {de-la Pe{\~n}a}, \citenamefont {Scheufele},
  \citenamefont {Grammer}, \citenamefont {Opel}, \citenamefont {Gepr{\"a}gs},
  \citenamefont {Cuevas}, \citenamefont {Gross}, \citenamefont {Huebl},
  \citenamefont {Kamra},\ and\ \citenamefont
  {Althammer}}]{guckelhorn2023observation}%
  \BibitemOpen
  \bibfield  {author} {\bibinfo {author} {\bibfnamefont {J.}~\bibnamefont
  {G{\"u}ckelhorn}}, \bibinfo {author} {\bibfnamefont {S.}~\bibnamefont {de-la
  Pe{\~n}a}}, \bibinfo {author} {\bibfnamefont {M.}~\bibnamefont {Scheufele}},
  \bibinfo {author} {\bibfnamefont {M.}~\bibnamefont {Grammer}}, \bibinfo
  {author} {\bibfnamefont {M.}~\bibnamefont {Opel}}, \bibinfo {author}
  {\bibfnamefont {S.}~\bibnamefont {Gepr{\"a}gs}}, \bibinfo {author}
  {\bibfnamefont {J.~C.}\ \bibnamefont {Cuevas}}, \bibinfo {author}
  {\bibfnamefont {R.}~\bibnamefont {Gross}}, \bibinfo {author} {\bibfnamefont
  {H.}~\bibnamefont {Huebl}}, \bibinfo {author} {\bibfnamefont
  {A.}~\bibnamefont {Kamra}},\ and\ \bibinfo {author} {\bibfnamefont
  {M.}~\bibnamefont {Althammer}},\ }\bibfield  {title} {\bibinfo {title}
  {Observation of the nonreciprocal magnon {Hanle} effect},\ }\href
  {https://doi.org/10.1103/PhysRevLett.130.216703} {\bibfield  {journal}
  {\bibinfo  {journal} {Phys. Rev. Lett.}\ }\textbf {\bibinfo {volume} {130}},\
  \bibinfo {pages} {216703} (\bibinfo {year} {2023})}\BibitemShut {NoStop}%
\bibitem [{\citenamefont {Kleinherbers}\ and\ \citenamefont
  {Tserkovnyak}(2024)}]{kleinherbers2024magnon}%
  \BibitemOpen
  \bibfield  {author} {\bibinfo {author} {\bibfnamefont {E.}~\bibnamefont
  {Kleinherbers}}\ and\ \bibinfo {author} {\bibfnamefont {Y.}~\bibnamefont
  {Tserkovnyak}},\ }\bibfield  {title} {\bibinfo {title} {Magnon kinetic theory
  of the antiferromagnetic {Hanle} effect},\ }\href
  {https://doi.org/10.1103/PhysRevB.110.L140408} {\bibfield  {journal}
  {\bibinfo  {journal} {Phys. Rev. B}\ }\textbf {\bibinfo {volume} {110}},\
  \bibinfo {pages} {L140408} (\bibinfo {year} {2024})}\BibitemShut {NoStop}%
\bibitem [{\citenamefont {Iguchi}\ \emph {et~al.}(2015)\citenamefont {Iguchi},
  \citenamefont {Uemura}, \citenamefont {Ueno},\ and\ \citenamefont
  {Onose}}]{iguchi2015nonreciprocal}%
  \BibitemOpen
  \bibfield  {author} {\bibinfo {author} {\bibfnamefont {Y.}~\bibnamefont
  {Iguchi}}, \bibinfo {author} {\bibfnamefont {S.}~\bibnamefont {Uemura}},
  \bibinfo {author} {\bibfnamefont {K.}~\bibnamefont {Ueno}},\ and\ \bibinfo
  {author} {\bibfnamefont {Y.}~\bibnamefont {Onose}},\ }\bibfield  {title}
  {\bibinfo {title} {Nonreciprocal magnon propagation in a noncentrosymmetric
  ferromagnet {LiFe}$_5${O}$_8$},\ }\href
  {https://doi.org/10.1103/PhysRevB.92.184419} {\bibfield  {journal} {\bibinfo
  {journal} {Phys. Rev. B}\ }\textbf {\bibinfo {volume} {92}},\ \bibinfo
  {pages} {184419} (\bibinfo {year} {2015})}\BibitemShut {NoStop}%
\bibitem [{\citenamefont {Pekar}\ and\ \citenamefont
  {Rashba}(1964)}]{pekar1964combined}%
  \BibitemOpen
  \bibfield  {author} {\bibinfo {author} {\bibfnamefont {S.}~\bibnamefont
  {Pekar}}\ and\ \bibinfo {author} {\bibfnamefont {E.}~\bibnamefont {Rashba}},\
  }\bibfield  {title} {\bibinfo {title} {Combined resonance in crystals in
  inhomogeneous magnetic fields},\ }\href@noop {} {\bibfield  {journal}
  {\bibinfo  {journal} {JETP}\ }\textbf {\bibinfo {volume} {47}} (\bibinfo
  {year} {1964})}\BibitemShut {NoStop}%
\bibitem [{\citenamefont {{\v{S}}mejkal}\ \emph {et~al.}(2020)\citenamefont
  {{\v{S}}mejkal}, \citenamefont {Gonz{\'a}lez-Hern{\'a}ndez}, \citenamefont
  {Jungwirth},\ and\ \citenamefont {Sinova}}]{vsmejkal2020crystal}%
  \BibitemOpen
  \bibfield  {author} {\bibinfo {author} {\bibfnamefont {L.}~\bibnamefont
  {{\v{S}}mejkal}}, \bibinfo {author} {\bibfnamefont {R.}~\bibnamefont
  {Gonz{\'a}lez-Hern{\'a}ndez}}, \bibinfo {author} {\bibfnamefont
  {T.}~\bibnamefont {Jungwirth}},\ and\ \bibinfo {author} {\bibfnamefont
  {J.}~\bibnamefont {Sinova}},\ }\bibfield  {title} {\bibinfo {title} {Crystal
  time-reversal symmetry breaking and spontaneous {Hall} effect in collinear
  antiferromagnets},\ }\href {https://doi.org/10.1126/sciadv.aaz8809}
  {\bibfield  {journal} {\bibinfo  {journal} {Sci. Adv.}\ }\textbf {\bibinfo
  {volume} {6}},\ \bibinfo {pages} {eaaz8809} (\bibinfo {year}
  {2020})}\BibitemShut {NoStop}%
\bibitem [{\citenamefont {Yuan}\ \emph {et~al.}(2020)\citenamefont {Yuan},
  \citenamefont {Wang}, \citenamefont {Luo}, \citenamefont {Rashba},\ and\
  \citenamefont {Zunger}}]{yuan2020giant}%
  \BibitemOpen
  \bibfield  {author} {\bibinfo {author} {\bibfnamefont {L.-D.}\ \bibnamefont
  {Yuan}}, \bibinfo {author} {\bibfnamefont {Z.}~\bibnamefont {Wang}}, \bibinfo
  {author} {\bibfnamefont {J.-W.}\ \bibnamefont {Luo}}, \bibinfo {author}
  {\bibfnamefont {E.~I.}\ \bibnamefont {Rashba}},\ and\ \bibinfo {author}
  {\bibfnamefont {A.}~\bibnamefont {Zunger}},\ }\bibfield  {title} {\bibinfo
  {title} {Giant momentum-dependent spin splitting in centrosymmetric low-{Z}
  antiferromagnets},\ }\href {https://doi.org/10.1103/PhysRevB.102.014422}
  {\bibfield  {journal} {\bibinfo  {journal} {Phys. Rev. B}\ }\textbf {\bibinfo
  {volume} {102}},\ \bibinfo {pages} {014422} (\bibinfo {year}
  {2020})}\BibitemShut {NoStop}%
\bibitem [{\citenamefont {Hayami}\ \emph {et~al.}(2020)\citenamefont {Hayami},
  \citenamefont {Yanagi},\ and\ \citenamefont {Kusunose}}]{hayami2020bottom}%
  \BibitemOpen
  \bibfield  {author} {\bibinfo {author} {\bibfnamefont {S.}~\bibnamefont
  {Hayami}}, \bibinfo {author} {\bibfnamefont {Y.}~\bibnamefont {Yanagi}},\
  and\ \bibinfo {author} {\bibfnamefont {H.}~\bibnamefont {Kusunose}},\
  }\bibfield  {title} {\bibinfo {title} {Bottom-up design of spin-split and
  reshaped electronic band structures in antiferromagnets without spin-orbit
  coupling: Procedure on the basis of augmented multipoles},\ }\href
  {https://doi.org/10.1103/PhysRevB.102.144441} {\bibfield  {journal} {\bibinfo
   {journal} {Phys. Rev. B}\ }\textbf {\bibinfo {volume} {102}},\ \bibinfo
  {pages} {144441} (\bibinfo {year} {2020})}\BibitemShut {NoStop}%
\bibitem [{\citenamefont {Yuan}\ \emph {et~al.}(2021)\citenamefont {Yuan},
  \citenamefont {Wang}, \citenamefont {Luo},\ and\ \citenamefont
  {Zunger}}]{yuan2021prediction}%
  \BibitemOpen
  \bibfield  {author} {\bibinfo {author} {\bibfnamefont {L.-D.}\ \bibnamefont
  {Yuan}}, \bibinfo {author} {\bibfnamefont {Z.}~\bibnamefont {Wang}}, \bibinfo
  {author} {\bibfnamefont {J.-W.}\ \bibnamefont {Luo}},\ and\ \bibinfo {author}
  {\bibfnamefont {A.}~\bibnamefont {Zunger}},\ }\bibfield  {title} {\bibinfo
  {title} {Prediction of low-{Z} collinear and noncollinear antiferromagnetic
  compounds having momentum-dependent spin splitting even without spin-orbit
  coupling},\ }\href {https://doi.org/10.1103/PhysRevMaterials.5.014409}
  {\bibfield  {journal} {\bibinfo  {journal} {Phys. Rev. Mater.}\ }\textbf
  {\bibinfo {volume} {5}},\ \bibinfo {pages} {014409} (\bibinfo {year}
  {2021})}\BibitemShut {NoStop}%
\bibitem [{\citenamefont {Krempask{\`y}}\ \emph {et~al.}(2024)\citenamefont
  {Krempask{\`y}}, \citenamefont {{\v{S}}mejkal}, \citenamefont {D'souza},
  \citenamefont {Hajlaoui}, \citenamefont {Springholz}, \citenamefont
  {Uhl{\'\i}{\v{r}}ov{\'a}}, \citenamefont {Alarab}, \citenamefont
  {Constantinou}, \citenamefont {Strocov}, \citenamefont {Usanov} \emph
  {et~al.}}]{krempasky2024altermagnetic}%
  \BibitemOpen
  \bibfield  {author} {\bibinfo {author} {\bibfnamefont {J.}~\bibnamefont
  {Krempask{\`y}}}, \bibinfo {author} {\bibfnamefont {L.}~\bibnamefont
  {{\v{S}}mejkal}}, \bibinfo {author} {\bibfnamefont {S.}~\bibnamefont
  {D'souza}}, \bibinfo {author} {\bibfnamefont {M.}~\bibnamefont {Hajlaoui}},
  \bibinfo {author} {\bibfnamefont {G.}~\bibnamefont {Springholz}}, \bibinfo
  {author} {\bibfnamefont {K.}~\bibnamefont {Uhl{\'\i}{\v{r}}ov{\'a}}},
  \bibinfo {author} {\bibfnamefont {F.}~\bibnamefont {Alarab}}, \bibinfo
  {author} {\bibfnamefont {P.}~\bibnamefont {Constantinou}}, \bibinfo {author}
  {\bibfnamefont {V.}~\bibnamefont {Strocov}}, \bibinfo {author} {\bibfnamefont
  {D.}~\bibnamefont {Usanov}}, \emph {et~al.},\ }\bibfield  {title} {\bibinfo
  {title} {Altermagnetic lifting of {Kramers} spin degeneracy},\ }\href
  {https://doi.org/10.1038/s41586-023-06907-7} {\bibfield  {journal} {\bibinfo
  {journal} {Nature}\ }\textbf {\bibinfo {volume} {626}},\ \bibinfo {pages}
  {517} (\bibinfo {year} {2024})}\BibitemShut {NoStop}%
\bibitem [{\citenamefont {Song}\ \emph {et~al.}(2025)\citenamefont {Song},
  \citenamefont {Bai}, \citenamefont {Zhou}, \citenamefont {Han}, \citenamefont
  {Reichlova}, \citenamefont {Dil}, \citenamefont {Liu}, \citenamefont {Chen},\
  and\ \citenamefont {Pan}}]{song2025altermagnets}%
  \BibitemOpen
  \bibfield  {author} {\bibinfo {author} {\bibfnamefont {C.}~\bibnamefont
  {Song}}, \bibinfo {author} {\bibfnamefont {H.}~\bibnamefont {Bai}}, \bibinfo
  {author} {\bibfnamefont {Z.}~\bibnamefont {Zhou}}, \bibinfo {author}
  {\bibfnamefont {L.}~\bibnamefont {Han}}, \bibinfo {author} {\bibfnamefont
  {H.}~\bibnamefont {Reichlova}}, \bibinfo {author} {\bibfnamefont {J.~H.}\
  \bibnamefont {Dil}}, \bibinfo {author} {\bibfnamefont {J.}~\bibnamefont
  {Liu}}, \bibinfo {author} {\bibfnamefont {X.}~\bibnamefont {Chen}},\ and\
  \bibinfo {author} {\bibfnamefont {F.}~\bibnamefont {Pan}},\ }\bibfield
  {title} {\bibinfo {title} {Altermagnets as a new class of functional
  materials},\ }\href {https://doi.org/10.1038/s41578-025-00779-1} {\bibfield
  {journal} {\bibinfo  {journal} {Nat. Rev. Mater.}\ }\textbf {\bibinfo
  {volume} {10}},\ \bibinfo {pages} {473} (\bibinfo {year} {2025})}\BibitemShut
  {NoStop}%
\bibitem [{\citenamefont {Duan}\ \emph {et~al.}(2025)\citenamefont {Duan},
  \citenamefont {Zhang}, \citenamefont {Zhu}, \citenamefont {Liu},
  \citenamefont {Zhang}, \citenamefont {\ifmmode \check{Z}\else
  \v{Z}\fi{}uti\ifmmode~\acute{c}\else \'{c}\fi{}},\ and\ \citenamefont
  {Zhou}}]{duan2025antiferroelectric}%
  \BibitemOpen
  \bibfield  {author} {\bibinfo {author} {\bibfnamefont {X.}~\bibnamefont
  {Duan}}, \bibinfo {author} {\bibfnamefont {J.}~\bibnamefont {Zhang}},
  \bibinfo {author} {\bibfnamefont {Z.}~\bibnamefont {Zhu}}, \bibinfo {author}
  {\bibfnamefont {Y.}~\bibnamefont {Liu}}, \bibinfo {author} {\bibfnamefont
  {Z.}~\bibnamefont {Zhang}}, \bibinfo {author} {\bibfnamefont
  {I.}~\bibnamefont {\ifmmode \check{Z}\else
  \v{Z}\fi{}uti\ifmmode~\acute{c}\else \'{c}\fi{}}},\ and\ \bibinfo {author}
  {\bibfnamefont {T.}~\bibnamefont {Zhou}},\ }\bibfield  {title} {\bibinfo
  {title} {Antiferroelectric altermagnets: Antiferroelectricity alters
  magnets},\ }\href {https://doi.org/10.1103/PhysRevLett.134.106801} {\bibfield
   {journal} {\bibinfo  {journal} {Phys. Rev. Lett.}\ }\textbf {\bibinfo
  {volume} {134}},\ \bibinfo {pages} {106801} (\bibinfo {year}
  {2025})}\BibitemShut {NoStop}%
\bibitem [{\citenamefont {Cao}\ \emph {et~al.}(2025)\citenamefont {Cao},
  \citenamefont {Denisov}, \citenamefont {Liu},\ and\ \citenamefont
  {{\v{Z}}uti{\'c}}}]{cao2025symmetry}%
  \BibitemOpen
  \bibfield  {author} {\bibinfo {author} {\bibfnamefont {J.~D.}\ \bibnamefont
  {Cao}}, \bibinfo {author} {\bibfnamefont {K.~S.}\ \bibnamefont {Denisov}},
  \bibinfo {author} {\bibfnamefont {Y.}~\bibnamefont {Liu}},\ and\ \bibinfo
  {author} {\bibfnamefont {I.}~\bibnamefont {{\v{Z}}uti{\'c}}},\ }\bibfield
  {title} {\bibinfo {title} {Symmetry classification for alternating excitons
  in two-dimensional altermagnets},\ }\href {https://doi.org/10.1103/zn7r-k1xd}
  {\bibfield  {journal} {\bibinfo  {journal} {Phys. Rev. Lett.}\ }\textbf
  {\bibinfo {volume} {135}},\ \bibinfo {pages} {266703} (\bibinfo {year}
  {2025})}\BibitemShut {NoStop}%
\bibitem [{\citenamefont {Zhu}\ \emph {et~al.}(2026)\citenamefont {Zhu},
  \citenamefont {Huang}, \citenamefont {Chen}, \citenamefont {Cui},
  \citenamefont {Duan}, \citenamefont {Zhang}, \citenamefont
  {{\v{Z}}uti{\'c}},\ and\ \citenamefont {Zhou}}]{zhu2026altermagnetic}%
  \BibitemOpen
  \bibfield  {author} {\bibinfo {author} {\bibfnamefont {Z.}~\bibnamefont
  {Zhu}}, \bibinfo {author} {\bibfnamefont {R.}~\bibnamefont {Huang}}, \bibinfo
  {author} {\bibfnamefont {X.}~\bibnamefont {Chen}}, \bibinfo {author}
  {\bibfnamefont {Z.}~\bibnamefont {Cui}}, \bibinfo {author} {\bibfnamefont
  {X.}~\bibnamefont {Duan}}, \bibinfo {author} {\bibfnamefont {J.}~\bibnamefont
  {Zhang}}, \bibinfo {author} {\bibfnamefont {I.}~\bibnamefont
  {{\v{Z}}uti{\'c}}},\ and\ \bibinfo {author} {\bibfnamefont {T.}~\bibnamefont
  {Zhou}},\ }\bibfield  {title} {\bibinfo {title} {Altermagnetic proximity
  effect},\ }\href {https://doi.org/10.1103/kqy8-myz1} {\bibfield  {journal}
  {\bibinfo  {journal} {Phys. Rev. Lett.}\ }\textbf {\bibinfo {volume} {136}},\
  \bibinfo {pages} {186702} (\bibinfo {year} {2026})}\BibitemShut {NoStop}%
\bibitem [{\citenamefont {Jungwirth}\ \emph {et~al.}(2026)\citenamefont
  {Jungwirth}, \citenamefont {Sinova}, \citenamefont {Fernandes}, \citenamefont
  {Liu}, \citenamefont {Watanabe}, \citenamefont {Murakami}, \citenamefont
  {Nakatsuji},\ and\ \citenamefont {{\v{S}}mejkal}}]{jungwirth2026symmetry}%
  \BibitemOpen
  \bibfield  {author} {\bibinfo {author} {\bibfnamefont {T.}~\bibnamefont
  {Jungwirth}}, \bibinfo {author} {\bibfnamefont {J.}~\bibnamefont {Sinova}},
  \bibinfo {author} {\bibfnamefont {R.~M.}\ \bibnamefont {Fernandes}}, \bibinfo
  {author} {\bibfnamefont {Q.}~\bibnamefont {Liu}}, \bibinfo {author}
  {\bibfnamefont {H.}~\bibnamefont {Watanabe}}, \bibinfo {author}
  {\bibfnamefont {S.}~\bibnamefont {Murakami}}, \bibinfo {author}
  {\bibfnamefont {S.}~\bibnamefont {Nakatsuji}},\ and\ \bibinfo {author}
  {\bibfnamefont {L.}~\bibnamefont {{\v{S}}mejkal}},\ }\bibfield  {title}
  {\bibinfo {title} {Symmetry, microscopy and spectroscopy signatures of
  altermagnetism},\ }\href {https://doi.org/10.1038/s41586-025-09883-2}
  {\bibfield  {journal} {\bibinfo  {journal} {Nature}\ }\textbf {\bibinfo
  {volume} {649}},\ \bibinfo {pages} {837} (\bibinfo {year}
  {2026})}\BibitemShut {NoStop}%
\bibitem [{\citenamefont {Brinkman}\ and\ \citenamefont
  {Elliott}(1966)}]{brinkman1966theory}%
  \BibitemOpen
  \bibfield  {author} {\bibinfo {author} {\bibfnamefont {W.}~\bibnamefont
  {Brinkman}}\ and\ \bibinfo {author} {\bibfnamefont {R.~J.}\ \bibnamefont
  {Elliott}},\ }\bibfield  {title} {\bibinfo {title} {Theory of spin-space
  groups},\ }\href {https://doi.org/10.1098/rspa.1966.0211} {\bibfield
  {journal} {\bibinfo  {journal} {Proc. R. Soc. Lond. A Math. Phys. Sci.}\
  }\textbf {\bibinfo {volume} {294}},\ \bibinfo {pages} {343} (\bibinfo {year}
  {1966})}\BibitemShut {NoStop}%
\bibitem [{\citenamefont {Litvin}\ and\ \citenamefont
  {Opechowski}(1974)}]{litvin1974spin}%
  \BibitemOpen
  \bibfield  {author} {\bibinfo {author} {\bibfnamefont {D.~B.}\ \bibnamefont
  {Litvin}}\ and\ \bibinfo {author} {\bibfnamefont {W.}~\bibnamefont
  {Opechowski}},\ }\bibfield  {title} {\bibinfo {title} {Spin groups},\ }\href
  {https://doi.org/10.1016/0031-8914(74)90157-8} {\bibfield  {journal}
  {\bibinfo  {journal} {Physica}\ }\textbf {\bibinfo {volume} {76}},\ \bibinfo
  {pages} {538} (\bibinfo {year} {1974})}\BibitemShut {NoStop}%
\bibitem [{\citenamefont {Liu}\ \emph {et~al.}(2022)\citenamefont {Liu},
  \citenamefont {Li}, \citenamefont {Han}, \citenamefont {Wan},\ and\
  \citenamefont {Liu}}]{liu2022spin}%
  \BibitemOpen
  \bibfield  {author} {\bibinfo {author} {\bibfnamefont {P.}~\bibnamefont
  {Liu}}, \bibinfo {author} {\bibfnamefont {J.}~\bibnamefont {Li}}, \bibinfo
  {author} {\bibfnamefont {J.}~\bibnamefont {Han}}, \bibinfo {author}
  {\bibfnamefont {X.}~\bibnamefont {Wan}},\ and\ \bibinfo {author}
  {\bibfnamefont {Q.}~\bibnamefont {Liu}},\ }\bibfield  {title} {\bibinfo
  {title} {Spin-group symmetry in magnetic materials with negligible spin-orbit
  coupling},\ }\href {https://doi.org/10.1103/PhysRevX.12.021016} {\bibfield
  {journal} {\bibinfo  {journal} {Phys. Rev. X}\ }\textbf {\bibinfo {volume}
  {12}},\ \bibinfo {pages} {021016} (\bibinfo {year} {2022})}\BibitemShut
  {NoStop}%
\bibitem [{\citenamefont {Chen}\ \emph {et~al.}(2024)\citenamefont {Chen},
  \citenamefont {Ren}, \citenamefont {Zhu}, \citenamefont {Yu}, \citenamefont
  {Zhang}, \citenamefont {Liu}, \citenamefont {Li}, \citenamefont {Liu},
  \citenamefont {Li},\ and\ \citenamefont {Liu}}]{chen2024enumeration}%
  \BibitemOpen
  \bibfield  {author} {\bibinfo {author} {\bibfnamefont {X.}~\bibnamefont
  {Chen}}, \bibinfo {author} {\bibfnamefont {J.}~\bibnamefont {Ren}}, \bibinfo
  {author} {\bibfnamefont {Y.}~\bibnamefont {Zhu}}, \bibinfo {author}
  {\bibfnamefont {Y.}~\bibnamefont {Yu}}, \bibinfo {author} {\bibfnamefont
  {A.}~\bibnamefont {Zhang}}, \bibinfo {author} {\bibfnamefont
  {P.}~\bibnamefont {Liu}}, \bibinfo {author} {\bibfnamefont {J.}~\bibnamefont
  {Li}}, \bibinfo {author} {\bibfnamefont {Y.}~\bibnamefont {Liu}}, \bibinfo
  {author} {\bibfnamefont {C.}~\bibnamefont {Li}},\ and\ \bibinfo {author}
  {\bibfnamefont {Q.}~\bibnamefont {Liu}},\ }\bibfield  {title} {\bibinfo
  {title} {{Enumeration} and {Representation} {Theory} of {Spin} {Space}
  {Groups}},\ }\href {https://doi.org/10.1103/PhysRevX.14.031038} {\bibfield
  {journal} {\bibinfo  {journal} {Phys. Rev. X}\ }\textbf {\bibinfo {volume}
  {14}},\ \bibinfo {pages} {031038} (\bibinfo {year} {2024})}\BibitemShut
  {NoStop}%
\bibitem [{\citenamefont {Liu}\ \emph {et~al.}(2026)\citenamefont {Liu},
  \citenamefont {Chen}, \citenamefont {Yu}, \citenamefont {Etxebarria},
  \citenamefont {Perez-Mato},\ and\ \citenamefont {Liu}}]{liu2026symmetry}%
  \BibitemOpen
  \bibfield  {author} {\bibinfo {author} {\bibfnamefont {Y.}~\bibnamefont
  {Liu}}, \bibinfo {author} {\bibfnamefont {X.}~\bibnamefont {Chen}}, \bibinfo
  {author} {\bibfnamefont {Y.}~\bibnamefont {Yu}}, \bibinfo {author}
  {\bibfnamefont {J.}~\bibnamefont {Etxebarria}}, \bibinfo {author}
  {\bibfnamefont {J.~M.}\ \bibnamefont {Perez-Mato}},\ and\ \bibinfo {author}
  {\bibfnamefont {Q.}~\bibnamefont {Liu}},\ }\bibfield  {title} {\bibinfo
  {title} {Symmetry classification of magnetic orders using oriented spin space
  groups},\ }\href {https://doi.org/10.1038/s41586-026-10401-1} {\bibfield
  {journal} {\bibinfo  {journal} {Nature}\ }\textbf {\bibinfo {volume} {652}},\
  \bibinfo {pages} {869} (\bibinfo {year} {2026})}\BibitemShut {NoStop}%
\bibitem [{\citenamefont {Cui}\ \emph {et~al.}(2023)\citenamefont {Cui},
  \citenamefont {Zeng}, \citenamefont {Cui}, \citenamefont {Yu},\ and\
  \citenamefont {Yang}}]{cui2023efficient}%
  \BibitemOpen
  \bibfield  {author} {\bibinfo {author} {\bibfnamefont {Q.}~\bibnamefont
  {Cui}}, \bibinfo {author} {\bibfnamefont {B.}~\bibnamefont {Zeng}}, \bibinfo
  {author} {\bibfnamefont {P.}~\bibnamefont {Cui}}, \bibinfo {author}
  {\bibfnamefont {T.}~\bibnamefont {Yu}},\ and\ \bibinfo {author}
  {\bibfnamefont {H.}~\bibnamefont {Yang}},\ }\bibfield  {title} {\bibinfo
  {title} {Efficient spin {Seebeck} and spin {Nernst} effects of magnons in
  altermagnets},\ }\href {https://doi.org/10.1103/PhysRevB.108.L180401}
  {\bibfield  {journal} {\bibinfo  {journal} {Phys. Rev. B}\ }\textbf {\bibinfo
  {volume} {108}},\ \bibinfo {pages} {L180401} (\bibinfo {year}
  {2023})}\BibitemShut {NoStop}%
\bibitem [{\citenamefont {Chen}\ \emph {et~al.}(2025)\citenamefont {Chen},
  \citenamefont {Liu}, \citenamefont {Liu}, \citenamefont {Yu}, \citenamefont
  {Ren}, \citenamefont {Li}, \citenamefont {Zhang},\ and\ \citenamefont
  {Liu}}]{chen2025unconventional}%
  \BibitemOpen
  \bibfield  {author} {\bibinfo {author} {\bibfnamefont {X.}~\bibnamefont
  {Chen}}, \bibinfo {author} {\bibfnamefont {Y.}~\bibnamefont {Liu}}, \bibinfo
  {author} {\bibfnamefont {P.}~\bibnamefont {Liu}}, \bibinfo {author}
  {\bibfnamefont {Y.}~\bibnamefont {Yu}}, \bibinfo {author} {\bibfnamefont
  {J.}~\bibnamefont {Ren}}, \bibinfo {author} {\bibfnamefont {J.}~\bibnamefont
  {Li}}, \bibinfo {author} {\bibfnamefont {A.}~\bibnamefont {Zhang}},\ and\
  \bibinfo {author} {\bibfnamefont {Q.}~\bibnamefont {Liu}},\ }\bibfield
  {title} {\bibinfo {title} {Unconventional magnons in collinear magnets
  dictated by spin space groups},\ }\bibfield  {journal} {\bibinfo  {journal}
  {Nature}\ }\textbf {\bibinfo {volume} {640}},\ \href
  {https://doi.org/10.1038/s41586-025-08715-7} {10.1038/s41586-025-08715-7}
  (\bibinfo {year} {2025})\BibitemShut {NoStop}%
\bibitem [{\citenamefont {S{\o}dequist}\ and\ \citenamefont
  {Olsen}(2024)}]{sodequist2024two}%
  \BibitemOpen
  \bibfield  {author} {\bibinfo {author} {\bibfnamefont {J.}~\bibnamefont
  {S{\o}dequist}}\ and\ \bibinfo {author} {\bibfnamefont {T.}~\bibnamefont
  {Olsen}},\ }\bibfield  {title} {\bibinfo {title} {Two-dimensional
  altermagnets from high throughput computational screening: {Symmetry}
  requirements, chiral magnons, and spin-orbit effects},\ }\bibfield  {journal}
  {\bibinfo  {journal} {Appl. Phys. Lett.}\ }\textbf {\bibinfo {volume}
  {124}},\ \href {https://doi.org/10.1063/5.0198285} {10.1063/5.0198285}
  (\bibinfo {year} {2024})\BibitemShut {NoStop}%
\bibitem [{\citenamefont {Liu}\ \emph {et~al.}(2024)\citenamefont {Liu},
  \citenamefont {Ozeki}, \citenamefont {Asai}, \citenamefont {Itoh},\ and\
  \citenamefont {Masuda}}]{liu2024chiral}%
  \BibitemOpen
  \bibfield  {author} {\bibinfo {author} {\bibfnamefont {Z.}~\bibnamefont
  {Liu}}, \bibinfo {author} {\bibfnamefont {M.}~\bibnamefont {Ozeki}}, \bibinfo
  {author} {\bibfnamefont {S.}~\bibnamefont {Asai}}, \bibinfo {author}
  {\bibfnamefont {S.}~\bibnamefont {Itoh}},\ and\ \bibinfo {author}
  {\bibfnamefont {T.}~\bibnamefont {Masuda}},\ }\bibfield  {title} {\bibinfo
  {title} {Chiral split magnon in altermagnetic {MnTe}},\ }\href
  {https://doi.org/10.1103/PhysRevLett.133.156702} {\bibfield  {journal}
  {\bibinfo  {journal} {Phys. Rev. Lett.}\ }\textbf {\bibinfo {volume} {133}},\
  \bibinfo {pages} {156702} (\bibinfo {year} {2024})}\BibitemShut {NoStop}%
\bibitem [{\citenamefont {Yu}\ \emph {et~al.}(2024)\citenamefont {Yu},
  \citenamefont {Feng}, \citenamefont {Zheng},\ and\ \citenamefont
  {Yao}}]{yu2024chiral}%
  \BibitemOpen
  \bibfield  {author} {\bibinfo {author} {\bibfnamefont {H.}~\bibnamefont
  {Yu}}, \bibinfo {author} {\bibfnamefont {W.}~\bibnamefont {Feng}}, \bibinfo
  {author} {\bibfnamefont {F.}~\bibnamefont {Zheng}},\ and\ \bibinfo {author}
  {\bibfnamefont {Y.}~\bibnamefont {Yao}},\ }\bibfield  {title} {\bibinfo
  {title} {Chiral magnons in an altermagnetic {Janus} {Mn$_2$SeTe} monolayer},\
  }\href {https://doi.org/10.1016/j.commt.2024.100021} {\bibfield  {journal}
  {\bibinfo  {journal} {Comp. Mat. Today}\ }\textbf {\bibinfo {volume} {4}},\
  \bibinfo {pages} {100021} (\bibinfo {year} {2024})}\BibitemShut {NoStop}%
\bibitem [{\citenamefont {Gomonay}\ \emph {et~al.}(2024)\citenamefont
  {Gomonay}, \citenamefont {Kravchuk}, \citenamefont {Jaeschke-Ubiergo},
  \citenamefont {Yershov}, \citenamefont {Jungwirth}, \citenamefont
  {{\v{S}}mejkal}, \citenamefont {Brink},\ and\ \citenamefont
  {Sinova}}]{gomonay2024structure}%
  \BibitemOpen
  \bibfield  {author} {\bibinfo {author} {\bibfnamefont {O.}~\bibnamefont
  {Gomonay}}, \bibinfo {author} {\bibfnamefont {V.~P.}\ \bibnamefont
  {Kravchuk}}, \bibinfo {author} {\bibfnamefont {R.}~\bibnamefont
  {Jaeschke-Ubiergo}}, \bibinfo {author} {\bibfnamefont {K.~V.}\ \bibnamefont
  {Yershov}}, \bibinfo {author} {\bibfnamefont {T.}~\bibnamefont {Jungwirth}},
  \bibinfo {author} {\bibfnamefont {L.}~\bibnamefont {{\v{S}}mejkal}}, \bibinfo
  {author} {\bibfnamefont {J.~v.~d.}\ \bibnamefont {Brink}},\ and\ \bibinfo
  {author} {\bibfnamefont {J.}~\bibnamefont {Sinova}},\ }\bibfield  {title}
  {\bibinfo {title} {Structure, control, and dynamics of altermagnetic
  textures},\ }\href {https://doi.org/10.1038/s44306-024-00042-3} {\bibfield
  {journal} {\bibinfo  {journal} {npj Spintronics}\ }\textbf {\bibinfo {volume}
  {2}},\ \bibinfo {pages} {35} (\bibinfo {year} {2024})}\BibitemShut {NoStop}%
\bibitem [{\citenamefont {Cui}\ \emph {et~al.}(2025)\citenamefont {Cui},
  \citenamefont {Bai},\ and\ \citenamefont {Delin}}]{cui2025anisotropic}%
  \BibitemOpen
  \bibfield  {author} {\bibinfo {author} {\bibfnamefont {Q.}~\bibnamefont
  {Cui}}, \bibinfo {author} {\bibfnamefont {X.}~\bibnamefont {Bai}},\ and\
  \bibinfo {author} {\bibfnamefont {A.}~\bibnamefont {Delin}},\ }\bibfield
  {title} {\bibinfo {title} {Anisotropic magnon transport in van der {Waals}
  ferromagnetic insulators},\ }\href {https://doi.org/10.1002/adfm.202407469}
  {\bibfield  {journal} {\bibinfo  {journal} {Adv. Funct. Mater.}\ }\textbf
  {\bibinfo {volume} {35}},\ \bibinfo {pages} {2407469} (\bibinfo {year}
  {2025})}\BibitemShut {NoStop}%
\bibitem [{\citenamefont {Sun}\ \emph {et~al.}(2025)\citenamefont {Sun},
  \citenamefont {Guo}, \citenamefont {Wang}, \citenamefont {Abernathy},
  \citenamefont {Tian},\ and\ \citenamefont {Li}}]{sun2025observation}%
  \BibitemOpen
  \bibfield  {author} {\bibinfo {author} {\bibfnamefont {Q.}~\bibnamefont
  {Sun}}, \bibinfo {author} {\bibfnamefont {J.}~\bibnamefont {Guo}}, \bibinfo
  {author} {\bibfnamefont {D.}~\bibnamefont {Wang}}, \bibinfo {author}
  {\bibfnamefont {D.~L.}\ \bibnamefont {Abernathy}}, \bibinfo {author}
  {\bibfnamefont {W.}~\bibnamefont {Tian}},\ and\ \bibinfo {author}
  {\bibfnamefont {C.}~\bibnamefont {Li}},\ }\bibfield  {title} {\bibinfo
  {title} {Observation of chiral magnon band splitting in altermagnetic
  hematite},\ }\href {https://doi.org/10.1103/7yhz-jptc} {\bibfield  {journal}
  {\bibinfo  {journal} {Phys. Rev. Lett.}\ }\textbf {\bibinfo {volume} {135}},\
  \bibinfo {pages} {186703} (\bibinfo {year} {2025})}\BibitemShut {NoStop}%
\bibitem [{\citenamefont {Sears}\ \emph {et~al.}(2026)\citenamefont {Sears},
  \citenamefont {Garlea}, \citenamefont {Lederman}, \citenamefont {Tranquada},\
  and\ \citenamefont {Zaliznyak}}]{sears2026altermagnetic}%
  \BibitemOpen
  \bibfield  {author} {\bibinfo {author} {\bibfnamefont {J.}~\bibnamefont
  {Sears}}, \bibinfo {author} {\bibfnamefont {V.~O.}\ \bibnamefont {Garlea}},
  \bibinfo {author} {\bibfnamefont {D.}~\bibnamefont {Lederman}}, \bibinfo
  {author} {\bibfnamefont {J.~M.}\ \bibnamefont {Tranquada}},\ and\ \bibinfo
  {author} {\bibfnamefont {I.~A.}\ \bibnamefont {Zaliznyak}},\ }\bibfield
  {title} {\bibinfo {title} {Altermagnetic and dipolar splitting of magnons in
  {FeF}$_2$},\ }\href {https://doi.org/10.1103/g6dt-rf8c} {\bibfield  {journal}
  {\bibinfo  {journal} {Phys. Rev. Lett.}\ }\textbf {\bibinfo {volume} {136}},\
  \bibinfo {pages} {226701} (\bibinfo {year} {2026})}\BibitemShut {NoStop}%
\bibitem [{\citenamefont {Jackson}(1998)}]{jackson1962classical}%
  \BibitemOpen
  \bibfield  {author} {\bibinfo {author} {\bibfnamefont {J.~D.}\ \bibnamefont
  {Jackson}},\ }\bibfield  {title} {\bibinfo {title} {Classical electrodynamics
  3rd ed.},\ }\href@noop {} {\bibfield  {journal} {\bibinfo  {journal} {Wiley,
  New York}\ } (\bibinfo {year} {1998})}\BibitemShut {NoStop}%
\bibitem [{\citenamefont {Flebus}\ and\ \citenamefont
  {Tserkovnyak}(2018)}]{flebus2018quantum}%
  \BibitemOpen
  \bibfield  {author} {\bibinfo {author} {\bibfnamefont {B.}~\bibnamefont
  {Flebus}}\ and\ \bibinfo {author} {\bibfnamefont {Y.}~\bibnamefont
  {Tserkovnyak}},\ }\bibfield  {title} {\bibinfo {title} {Quantum-impurity
  relaxometry of magnetization dynamics},\ }\href
  {https://doi.org/10.1103/PhysRevLett.121.187204} {\bibfield  {journal}
  {\bibinfo  {journal} {Phys. Rev. Lett.}\ }\textbf {\bibinfo {volume} {121}},\
  \bibinfo {pages} {187204} (\bibinfo {year} {2018})}\BibitemShut {NoStop}%
\bibitem [{\citenamefont {Wang}\ \emph {et~al.}(2022)\citenamefont {Wang},
  \citenamefont {Zhang}, \citenamefont {McLaughlin}, \citenamefont {Flebus},
  \citenamefont {Huang}, \citenamefont {Xiao}, \citenamefont {Liu},
  \citenamefont {Wu}, \citenamefont {Fullerton},\ and\ \citenamefont
  {Tserkovnyak}}]{wang2022noninvasive}%
  \BibitemOpen
  \bibfield  {author} {\bibinfo {author} {\bibfnamefont {H.}~\bibnamefont
  {Wang}}, \bibinfo {author} {\bibfnamefont {S.}~\bibnamefont {Zhang}},
  \bibinfo {author} {\bibfnamefont {N.~J.}\ \bibnamefont {McLaughlin}},
  \bibinfo {author} {\bibfnamefont {B.}~\bibnamefont {Flebus}}, \bibinfo
  {author} {\bibfnamefont {M.}~\bibnamefont {Huang}}, \bibinfo {author}
  {\bibfnamefont {D.}~\bibnamefont {Xiao}}, \bibinfo {author} {\bibfnamefont
  {C.}~\bibnamefont {Liu}}, \bibinfo {author} {\bibfnamefont {M.}~\bibnamefont
  {Wu}}, \bibinfo {author} {\bibfnamefont {E.~E.}\ \bibnamefont {Fullerton}},\
  and\ \bibinfo {author} {\bibfnamefont {Y.}~\bibnamefont {Tserkovnyak}},\
  }\bibfield  {title} {\bibinfo {title} {Noninvasive measurements of spin
  transport properties of an antiferromagnetic insulator},\ }\href
  {https://doi.org/10.1126/sciadv.abg8562} {\bibfield  {journal} {\bibinfo
  {journal} {Sci. Advances}\ }\textbf {\bibinfo {volume} {8}},\ \bibinfo
  {pages} {eabg8562} (\bibinfo {year} {2022})}\BibitemShut {NoStop}%
\bibitem [{\citenamefont {Heitzer}\ \emph {et~al.}(2024)\citenamefont
  {Heitzer}, \citenamefont {Pinto}, \citenamefont {Rodr{\'\i}guez},
  \citenamefont {Rodr{\'\i}guez-Su{\'a}rez},\ and\ \citenamefont
  {Maze}}]{heitzer2024characterization}%
  \BibitemOpen
  \bibfield  {author} {\bibinfo {author} {\bibfnamefont {R.}~\bibnamefont
  {Heitzer}}, \bibinfo {author} {\bibfnamefont {F.}~\bibnamefont {Pinto}},
  \bibinfo {author} {\bibfnamefont {E.}~\bibnamefont {Rodr{\'\i}guez}},
  \bibinfo {author} {\bibfnamefont {R.}~\bibnamefont
  {Rodr{\'\i}guez-Su{\'a}rez}},\ and\ \bibinfo {author} {\bibfnamefont
  {J.}~\bibnamefont {Maze}},\ }\bibfield  {title} {\bibinfo {title}
  {Characterization of antiferromagnetic magnons using nitrogen vacancy center
  relaxometry},\ }\href {https://doi.org/10.1103/PhysRevB.110.134431}
  {\bibfield  {journal} {\bibinfo  {journal} {Phys. Rev. B}\ }\textbf {\bibinfo
  {volume} {110}},\ \bibinfo {pages} {134431} (\bibinfo {year}
  {2024})}\BibitemShut {NoStop}%
\bibitem [{\citenamefont {Melendez}\ \emph {et~al.}(2025)\citenamefont
  {Melendez}, \citenamefont {Das}, \citenamefont {Rodriguez}, \citenamefont
  {Kao}, \citenamefont {Liu}, \citenamefont {Williams}, \citenamefont {Lv},
  \citenamefont {Goldberger}, \citenamefont {Chatterjee},\ and\ \citenamefont
  {Singh}}]{melendez2025quantum}%
  \BibitemOpen
  \bibfield  {author} {\bibinfo {author} {\bibfnamefont {A.~L.}\ \bibnamefont
  {Melendez}}, \bibinfo {author} {\bibfnamefont {S.}~\bibnamefont {Das}},
  \bibinfo {author} {\bibfnamefont {F.~A.}\ \bibnamefont {Rodriguez}}, \bibinfo
  {author} {\bibfnamefont {I.-H.}\ \bibnamefont {Kao}}, \bibinfo {author}
  {\bibfnamefont {W.}~\bibnamefont {Liu}}, \bibinfo {author} {\bibfnamefont
  {A.~J.}\ \bibnamefont {Williams}}, \bibinfo {author} {\bibfnamefont
  {B.}~\bibnamefont {Lv}}, \bibinfo {author} {\bibfnamefont {J.}~\bibnamefont
  {Goldberger}}, \bibinfo {author} {\bibfnamefont {S.}~\bibnamefont
  {Chatterjee}},\ and\ \bibinfo {author} {\bibfnamefont {S.}~\bibnamefont
  {Singh}},\ }\bibfield  {title} {\bibinfo {title} {Quantum sensing of
  broadband spin dynamics and magnon transport in antiferromagnets},\ }\href
  {https://doi.org/10.1126/sciadv.adu9381} {\bibfield  {journal} {\bibinfo
  {journal} {Sci. Adv.}\ }\textbf {\bibinfo {volume} {11}},\ \bibinfo {pages}
  {eadu9381} (\bibinfo {year} {2025})}\BibitemShut {NoStop}%
\bibitem [{\citenamefont {Bittencourt}\ \emph {et~al.}(2026)\citenamefont
  {Bittencourt}, \citenamefont {Hosseinabadi}, \citenamefont {Sinova},
  \citenamefont {{\v{S}}mejkal},\ and\ \citenamefont
  {Marino}}]{bittencourt2026quantum}%
  \BibitemOpen
  \bibfield  {author} {\bibinfo {author} {\bibfnamefont {V.}~\bibnamefont
  {Bittencourt}}, \bibinfo {author} {\bibfnamefont {H.}~\bibnamefont
  {Hosseinabadi}}, \bibinfo {author} {\bibfnamefont {J.}~\bibnamefont
  {Sinova}}, \bibinfo {author} {\bibfnamefont {L.}~\bibnamefont
  {{\v{S}}mejkal}},\ and\ \bibinfo {author} {\bibfnamefont {J.}~\bibnamefont
  {Marino}},\ }\bibfield  {title} {\bibinfo {title} {Quantum {Impurity}
  {Sensing} of {Altermagnetic} {Order}},\ }\href
  {https://doi.org/10.1103/2ppn-kvjv} {\bibfield  {journal} {\bibinfo
  {journal} {Phys. Rev. Lett.}\ }\textbf {\bibinfo {volume} {136}},\ \bibinfo
  {pages} {146701} (\bibinfo {year} {2026})}\BibitemShut {NoStop}%
\bibitem [{\citenamefont {Lebrun}\ \emph {et~al.}(2018)\citenamefont {Lebrun},
  \citenamefont {Ross}, \citenamefont {Bender}, \citenamefont {Qaiumzadeh},
  \citenamefont {Baldrati}, \citenamefont {Cramer}, \citenamefont {Brataas},
  \citenamefont {Duine},\ and\ \citenamefont {Kl{\"a}ui}}]{lebrun2018tunable}%
  \BibitemOpen
  \bibfield  {author} {\bibinfo {author} {\bibfnamefont {R.}~\bibnamefont
  {Lebrun}}, \bibinfo {author} {\bibfnamefont {A.}~\bibnamefont {Ross}},
  \bibinfo {author} {\bibfnamefont {S.~A.}\ \bibnamefont {Bender}}, \bibinfo
  {author} {\bibfnamefont {A.}~\bibnamefont {Qaiumzadeh}}, \bibinfo {author}
  {\bibfnamefont {L.}~\bibnamefont {Baldrati}}, \bibinfo {author}
  {\bibfnamefont {J.}~\bibnamefont {Cramer}}, \bibinfo {author} {\bibfnamefont
  {A.}~\bibnamefont {Brataas}}, \bibinfo {author} {\bibfnamefont {R.~A.}\
  \bibnamefont {Duine}},\ and\ \bibinfo {author} {\bibfnamefont
  {M.}~\bibnamefont {Kl{\"a}ui}},\ }\bibfield  {title} {\bibinfo {title}
  {Tunable long-distance spin transport in a crystalline antiferromagnetic iron
  oxide},\ }\href {https://doi.org/10.1038/s41586-018-0490-7} {\bibfield
  {journal} {\bibinfo  {journal} {Nature}\ }\textbf {\bibinfo {volume} {561}},\
  \bibinfo {pages} {222} (\bibinfo {year} {2018})}\BibitemShut {NoStop}%
\bibitem [{\citenamefont {Ross}\ \emph {et~al.}(2019)\citenamefont {Ross},
  \citenamefont {Lebrun}, \citenamefont {Gomonay}, \citenamefont {Grave},
  \citenamefont {Kay}, \citenamefont {Baldrati}, \citenamefont {Becker},
  \citenamefont {Qaiumzadeh}, \citenamefont {Ulloa},\ and\ \citenamefont
  {Jakob}}]{ross2019propagation}%
  \BibitemOpen
  \bibfield  {author} {\bibinfo {author} {\bibfnamefont {A.}~\bibnamefont
  {Ross}}, \bibinfo {author} {\bibfnamefont {R.}~\bibnamefont {Lebrun}},
  \bibinfo {author} {\bibfnamefont {O.}~\bibnamefont {Gomonay}}, \bibinfo
  {author} {\bibfnamefont {D.~A.}\ \bibnamefont {Grave}}, \bibinfo {author}
  {\bibfnamefont {A.}~\bibnamefont {Kay}}, \bibinfo {author} {\bibfnamefont
  {L.}~\bibnamefont {Baldrati}}, \bibinfo {author} {\bibfnamefont
  {S.}~\bibnamefont {Becker}}, \bibinfo {author} {\bibfnamefont
  {A.}~\bibnamefont {Qaiumzadeh}}, \bibinfo {author} {\bibfnamefont
  {C.}~\bibnamefont {Ulloa}},\ and\ \bibinfo {author} {\bibfnamefont
  {G.}~\bibnamefont {Jakob}},\ }\bibfield  {title} {\bibinfo {title}
  {Propagation length of antiferromagnetic magnons governed by domain
  configurations},\ }\href {https://doi.org/10.1021/acs.nanolett.9b03837}
  {\bibfield  {journal} {\bibinfo  {journal} {Nano Lett.}\ }\textbf {\bibinfo
  {volume} {20}},\ \bibinfo {pages} {306} (\bibinfo {year} {2019})}\BibitemShut
  {NoStop}%
\bibitem [{\citenamefont {Troncoso}\ \emph {et~al.}(2020)\citenamefont
  {Troncoso}, \citenamefont {Bender}, \citenamefont {Brataas},\ and\
  \citenamefont {Duine}}]{troncoso2020spin}%
  \BibitemOpen
  \bibfield  {author} {\bibinfo {author} {\bibfnamefont {R.~E.}\ \bibnamefont
  {Troncoso}}, \bibinfo {author} {\bibfnamefont {S.~A.}\ \bibnamefont
  {Bender}}, \bibinfo {author} {\bibfnamefont {A.}~\bibnamefont {Brataas}},\
  and\ \bibinfo {author} {\bibfnamefont {R.~A.}\ \bibnamefont {Duine}},\
  }\bibfield  {title} {\bibinfo {title} {Spin transport in thick insulating
  antiferromagnetic films},\ }\href
  {https://doi.org/10.1103/PhysRevB.101.054404} {\bibfield  {journal} {\bibinfo
   {journal} {Phys. Rev. B}\ }\textbf {\bibinfo {volume} {101}},\ \bibinfo
  {pages} {054404} (\bibinfo {year} {2020})}\BibitemShut {NoStop}%
\bibitem [{\citenamefont {Van~Rijn}\ \emph {et~al.}(2024)\citenamefont
  {Van~Rijn}, \citenamefont {Van~Wees},\ and\ \citenamefont
  {Banerjee}}]{van2024magnon}%
  \BibitemOpen
  \bibfield  {author} {\bibinfo {author} {\bibfnamefont {J.}~\bibnamefont
  {Van~Rijn}}, \bibinfo {author} {\bibfnamefont {B.~J.}\ \bibnamefont
  {Van~Wees}},\ and\ \bibinfo {author} {\bibfnamefont {T.}~\bibnamefont
  {Banerjee}},\ }\bibfield  {title} {\bibinfo {title} {Magnon transport in
  antiferromagnetic strained {SrMnO}$_3$ thin films},\ }\href
  {https://doi.org/10.1103/PhysRevB.110.144446} {\bibfield  {journal} {\bibinfo
   {journal} {Phys. Rev. B}\ }\textbf {\bibinfo {volume} {110}},\ \bibinfo
  {pages} {144446} (\bibinfo {year} {2024})}\BibitemShut {NoStop}%
\bibitem [{\citenamefont {Olsson}\ \emph {et~al.}(2020)\citenamefont {Olsson},
  \citenamefont {An}, \citenamefont {Fiete}, \citenamefont {Zhou},
  \citenamefont {Shi},\ and\ \citenamefont {Li}}]{olsson2020pure}%
  \BibitemOpen
  \bibfield  {author} {\bibinfo {author} {\bibfnamefont {K.~S.}\ \bibnamefont
  {Olsson}}, \bibinfo {author} {\bibfnamefont {K.}~\bibnamefont {An}}, \bibinfo
  {author} {\bibfnamefont {G.~A.}\ \bibnamefont {Fiete}}, \bibinfo {author}
  {\bibfnamefont {J.}~\bibnamefont {Zhou}}, \bibinfo {author} {\bibfnamefont
  {L.}~\bibnamefont {Shi}},\ and\ \bibinfo {author} {\bibfnamefont
  {X.}~\bibnamefont {Li}},\ }\bibfield  {title} {\bibinfo {title} {Pure spin
  current and magnon chemical potential in a nonequilibrium magnetic
  insulator},\ }\href {https://doi.org/10.1103/PhysRevX.10.021029} {\bibfield
  {journal} {\bibinfo  {journal} {Phys. Rev. X}\ }\textbf {\bibinfo {volume}
  {10}},\ \bibinfo {pages} {021029} (\bibinfo {year} {2020})}\BibitemShut
  {NoStop}%
\bibitem [{\citenamefont {Kargar}\ and\ \citenamefont
  {Balandin}(2021)}]{kargar2021advances}%
  \BibitemOpen
  \bibfield  {author} {\bibinfo {author} {\bibfnamefont {F.}~\bibnamefont
  {Kargar}}\ and\ \bibinfo {author} {\bibfnamefont {A.~A.}\ \bibnamefont
  {Balandin}},\ }\bibfield  {title} {\bibinfo {title} {Advances in
  {Brillouin}--{Mandelstam} light-scattering spectroscopy},\ }\href
  {https://doi.org/10.1038/s41566-021-00836-5} {\bibfield  {journal} {\bibinfo
  {journal} {Nat. Photon.}\ }\textbf {\bibinfo {volume} {15}},\ \bibinfo
  {pages} {720} (\bibinfo {year} {2021})}\BibitemShut {NoStop}%
\bibitem [{\citenamefont {Gu}\ \emph {et~al.}(2025)\citenamefont {Gu},
  \citenamefont {Barker}, \citenamefont {Li}, \citenamefont {Kikkawa},
  \citenamefont {Camino}, \citenamefont {Kisslinger}, \citenamefont
  {Sinsheimer}, \citenamefont {Lienhard}, \citenamefont {Bauer}, \citenamefont
  {Ross} \emph {et~al.}}]{gu2025observing}%
  \BibitemOpen
  \bibfield  {author} {\bibinfo {author} {\bibfnamefont {Y.}~\bibnamefont
  {Gu}}, \bibinfo {author} {\bibfnamefont {J.}~\bibnamefont {Barker}}, \bibinfo
  {author} {\bibfnamefont {J.}~\bibnamefont {Li}}, \bibinfo {author}
  {\bibfnamefont {T.}~\bibnamefont {Kikkawa}}, \bibinfo {author} {\bibfnamefont
  {F.}~\bibnamefont {Camino}}, \bibinfo {author} {\bibfnamefont
  {K.}~\bibnamefont {Kisslinger}}, \bibinfo {author} {\bibfnamefont
  {J.}~\bibnamefont {Sinsheimer}}, \bibinfo {author} {\bibfnamefont
  {L.}~\bibnamefont {Lienhard}}, \bibinfo {author} {\bibfnamefont {J.~J.}\
  \bibnamefont {Bauer}}, \bibinfo {author} {\bibfnamefont {C.~A.}\ \bibnamefont
  {Ross}}, \emph {et~al.},\ }\bibfield  {title} {\bibinfo {title} {Observing
  differential spin currents by resonant inelastic {X}-ray scattering},\ }\href
  {https://doi.org/10.1038/s41586-025-09488-9} {\bibfield  {journal} {\bibinfo
  {journal} {Nature}\ }\textbf {\bibinfo {volume} {645}},\ \bibinfo {pages}
  {900} (\bibinfo {year} {2025})}\BibitemShut {NoStop}%
\bibitem [{\citenamefont {Wittrock}\ \emph {et~al.}(2026)\citenamefont
  {Wittrock}, \citenamefont {Klose}, \citenamefont {Perna}, \citenamefont
  {Baumgaertl}, \citenamefont {Mucchietto}, \citenamefont {Schneider},
  \citenamefont {Fuchs}, \citenamefont {Deinhart}, \citenamefont {Karaman},
  \citenamefont {Grundler} \emph {et~al.}}]{wittrock2026soft}%
  \BibitemOpen
  \bibfield  {author} {\bibinfo {author} {\bibfnamefont {S.}~\bibnamefont
  {Wittrock}}, \bibinfo {author} {\bibfnamefont {C.}~\bibnamefont {Klose}},
  \bibinfo {author} {\bibfnamefont {S.}~\bibnamefont {Perna}}, \bibinfo
  {author} {\bibfnamefont {K.}~\bibnamefont {Baumgaertl}}, \bibinfo {author}
  {\bibfnamefont {A.}~\bibnamefont {Mucchietto}}, \bibinfo {author}
  {\bibfnamefont {M.}~\bibnamefont {Schneider}}, \bibinfo {author}
  {\bibfnamefont {J.}~\bibnamefont {Fuchs}}, \bibinfo {author} {\bibfnamefont
  {V.}~\bibnamefont {Deinhart}}, \bibinfo {author} {\bibfnamefont
  {T.}~\bibnamefont {Karaman}}, \bibinfo {author} {\bibfnamefont
  {D.}~\bibnamefont {Grundler}}, \emph {et~al.},\ }\bibfield  {title} {\bibinfo
  {title} {Soft-{X}-ray momentum microscopy of nonlinear magnon interactions},\
  }\href {https://doi.org/10.1038/s41567-026-03318-z} {\bibfield  {journal}
  {\bibinfo  {journal} {Nat. Phys.}\ }\textbf {\bibinfo {volume} {22}},\
  \bibinfo {pages} {1160} (\bibinfo {year} {2026})}\BibitemShut {NoStop}%
\bibitem [{\citenamefont {Karimeddiny}\ \emph {et~al.}(2023)\citenamefont
  {Karimeddiny}, \citenamefont {Cham}, \citenamefont {Smedley}, \citenamefont
  {Ralph},\ and\ \citenamefont {Luo}}]{karimeddiny2023sagnac}%
  \BibitemOpen
  \bibfield  {author} {\bibinfo {author} {\bibfnamefont {S.}~\bibnamefont
  {Karimeddiny}}, \bibinfo {author} {\bibfnamefont {T.~M.~J.}\ \bibnamefont
  {Cham}}, \bibinfo {author} {\bibfnamefont {O.}~\bibnamefont {Smedley}},
  \bibinfo {author} {\bibfnamefont {D.~C.}\ \bibnamefont {Ralph}},\ and\
  \bibinfo {author} {\bibfnamefont {Y.~K.}\ \bibnamefont {Luo}},\ }\bibfield
  {title} {\bibinfo {title} {Sagnac interferometry for high-sensitivity optical
  measurements of spin-orbit torque},\ }\href
  {https://doi.org/10.1126/sciadv.adi9039} {\bibfield  {journal} {\bibinfo
  {journal} {Sci. adv.}\ }\textbf {\bibinfo {volume} {9}},\ \bibinfo {pages}
  {eadi9039} (\bibinfo {year} {2023})}\BibitemShut {NoStop}%
\bibitem [{\citenamefont {Dyakonov}\ and\ \citenamefont
  {Perel}(1972)}]{dyakonov1972spin}%
  \BibitemOpen
  \bibfield  {author} {\bibinfo {author} {\bibfnamefont {M.}~\bibnamefont
  {Dyakonov}}\ and\ \bibinfo {author} {\bibfnamefont {V.}~\bibnamefont
  {Perel}},\ }\bibfield  {title} {\bibinfo {title} {Spin relaxation of
  conduction electrons in noncentrosymmetric semiconductors},\ }\href@noop {}
  {\bibfield  {journal} {\bibinfo  {journal} {Sov. Phys. Solid State}\ }\textbf
  {\bibinfo {volume} {13}},\ \bibinfo {pages} {3023} (\bibinfo {year}
  {1972})}\BibitemShut {NoStop}%
\bibitem [{SM()}]{SM}%
  \BibitemOpen
  \href@noop {} {\bibinfo {title} {{See Supplemental Material for expanded
  discussion of theoretical and computational methods.}}}\BibitemShut {Stop}%
\bibitem [{\citenamefont {Kawano}\ and\ \citenamefont
  {Hotta}(2019{\natexlab{b}})}]{kawano2019thermal}%
  \BibitemOpen
  \bibfield  {author} {\bibinfo {author} {\bibfnamefont {M.}~\bibnamefont
  {Kawano}}\ and\ \bibinfo {author} {\bibfnamefont {C.}~\bibnamefont {Hotta}},\
  }\bibfield  {title} {\bibinfo {title} {Thermal {Hall} effect and topological
  edge states in a square-lattice antiferromagnet},\ }\href
  {https://doi.org/10.1103/PhysRevB.99.054422} {\bibfield  {journal} {\bibinfo
  {journal} {Phys. Rev. B}\ }\textbf {\bibinfo {volume} {99}},\ \bibinfo
  {pages} {054422} (\bibinfo {year} {2019}{\natexlab{b}})}\BibitemShut
  {NoStop}%
\bibitem [{\citenamefont {White}\ \emph {et~al.}(1965)\citenamefont {White},
  \citenamefont {Sparks},\ and\ \citenamefont
  {Ortenburger}}]{white1965diagonalization}%
  \BibitemOpen
  \bibfield  {author} {\bibinfo {author} {\bibfnamefont {R.~M.}\ \bibnamefont
  {White}}, \bibinfo {author} {\bibfnamefont {M.}~\bibnamefont {Sparks}},\ and\
  \bibinfo {author} {\bibfnamefont {I.}~\bibnamefont {Ortenburger}},\
  }\bibfield  {title} {\bibinfo {title} {Diagonalization of the
  antiferromagnetic magnon-phonon interaction},\ }\href
  {https://doi.org/10.1103/PhysRev.139.A450} {\bibfield  {journal} {\bibinfo
  {journal} {Phys. Rev.}\ }\textbf {\bibinfo {volume} {139}},\ \bibinfo {pages}
  {A450} (\bibinfo {year} {1965})}\BibitemShut {NoStop}%
\bibitem [{\citenamefont {Colpa}(1978)}]{colpa1978diagonalization}%
  \BibitemOpen
  \bibfield  {author} {\bibinfo {author} {\bibfnamefont {J.~H.~P.}\
  \bibnamefont {Colpa}},\ }\bibfield  {title} {\bibinfo {title}
  {Diagonalization of the quadratic boson hamiltonian},\ }\href
  {https://doi.org/10.1016/0378-4371(78)90160-7} {\bibfield  {journal}
  {\bibinfo  {journal} {Physica A: Statistical Mechanics and its Applications}\
  }\textbf {\bibinfo {volume} {93}},\ \bibinfo {pages} {327} (\bibinfo {year}
  {1978})}\BibitemShut {NoStop}%
\bibitem [{\citenamefont {Rezende}\ \emph {et~al.}(2019)\citenamefont
  {Rezende}, \citenamefont {Azevedo},\ and\ \citenamefont
  {Rodr{\'\i}guez-Su{\'a}rez}}]{rezende2019introduction}%
  \BibitemOpen
  \bibfield  {author} {\bibinfo {author} {\bibfnamefont {S.~M.}\ \bibnamefont
  {Rezende}}, \bibinfo {author} {\bibfnamefont {A.}~\bibnamefont {Azevedo}},\
  and\ \bibinfo {author} {\bibfnamefont {R.~L.}\ \bibnamefont
  {Rodr{\'\i}guez-Su{\'a}rez}},\ }\bibfield  {title} {\bibinfo {title}
  {Introduction to antiferromagnetic magnons},\ }\bibfield  {journal} {\bibinfo
   {journal} {J. Appl. Phys.}\ }\textbf {\bibinfo {volume} {126}},\ \href
  {https://doi.org/10.1063/1.5109132} {10.1063/1.5109132} (\bibinfo {year}
  {2019})\BibitemShut {NoStop}%
\bibitem [{\citenamefont {Nambu}\ \emph {et~al.}(2020)\citenamefont {Nambu},
  \citenamefont {Barker}, \citenamefont {Okino}, \citenamefont {Kikkawa},
  \citenamefont {Shiomi}, \citenamefont {Enderle}, \citenamefont {Weber},
  \citenamefont {Winn}, \citenamefont {Graves-Brook},\ and\ \citenamefont
  {Tranquada}}]{nambu2020observation}%
  \BibitemOpen
  \bibfield  {author} {\bibinfo {author} {\bibfnamefont {Y.}~\bibnamefont
  {Nambu}}, \bibinfo {author} {\bibfnamefont {J.}~\bibnamefont {Barker}},
  \bibinfo {author} {\bibfnamefont {Y.}~\bibnamefont {Okino}}, \bibinfo
  {author} {\bibfnamefont {T.}~\bibnamefont {Kikkawa}}, \bibinfo {author}
  {\bibfnamefont {Y.}~\bibnamefont {Shiomi}}, \bibinfo {author} {\bibfnamefont
  {M.}~\bibnamefont {Enderle}}, \bibinfo {author} {\bibfnamefont
  {T.}~\bibnamefont {Weber}}, \bibinfo {author} {\bibfnamefont
  {B.}~\bibnamefont {Winn}}, \bibinfo {author} {\bibfnamefont {M.}~\bibnamefont
  {Graves-Brook}},\ and\ \bibinfo {author} {\bibfnamefont {J.~M.}\ \bibnamefont
  {Tranquada}},\ }\bibfield  {title} {\bibinfo {title} {Observation of magnon
  polarization},\ }\href {https://doi.org/10.1103/PhysRevLett.125.027201}
  {\bibfield  {journal} {\bibinfo  {journal} {Phys. Rev. Lett.}\ }\textbf
  {\bibinfo {volume} {125}},\ \bibinfo {pages} {027201} (\bibinfo {year}
  {2020})}\BibitemShut {NoStop}%
\bibitem [{\citenamefont {Dyakonov}(2017)}]{dyakonov2017spin}%
  \BibitemOpen
  \bibfield  {author} {\bibinfo {author} {\bibfnamefont {M.~I.}\ \bibnamefont
  {Dyakonov}},\ }\href@noop {} {\emph {\bibinfo {title} {Spin physics in
  semiconductors}}},\ Vol.~\bibinfo {volume} {1}\ (\bibinfo  {publisher}
  {Springer},\ \bibinfo {year} {2017})\BibitemShut {NoStop}%
\bibitem [{\citenamefont {Cheng}\ \emph {et~al.}(2016)\citenamefont {Cheng},
  \citenamefont {Daniels}, \citenamefont {Zhu},\ and\ \citenamefont
  {Xiao}}]{cheng2016antiferromagnetic}%
  \BibitemOpen
  \bibfield  {author} {\bibinfo {author} {\bibfnamefont {R.}~\bibnamefont
  {Cheng}}, \bibinfo {author} {\bibfnamefont {M.~W.}\ \bibnamefont {Daniels}},
  \bibinfo {author} {\bibfnamefont {J.-G.}\ \bibnamefont {Zhu}},\ and\ \bibinfo
  {author} {\bibfnamefont {D.}~\bibnamefont {Xiao}},\ }\bibfield  {title}
  {\bibinfo {title} {Antiferromagnetic spin wave field-effect transistor},\
  }\href {https://doi.org/10.1038/srep24223} {\bibfield  {journal} {\bibinfo
  {journal} {Sci. Rep.}\ }\textbf {\bibinfo {volume} {6}},\ \bibinfo {pages}
  {24223} (\bibinfo {year} {2016})}\BibitemShut {NoStop}%
\bibitem [{\citenamefont {Landau}\ \emph {et~al.}(1975)\citenamefont {Landau},
  \citenamefont {Lifshitz},\ and\ \citenamefont
  {Lifshitz}}]{landau1975classical}%
  \BibitemOpen
  \bibfield  {author} {\bibinfo {author} {\bibfnamefont {L.~D.}\ \bibnamefont
  {Landau}}, \bibinfo {author} {\bibfnamefont {E.~M.}\ \bibnamefont
  {Lifshitz}},\ and\ \bibinfo {author} {\bibfnamefont {E.~M.}\ \bibnamefont
  {Lifshitz}},\ }\href@noop {} {\emph {\bibinfo {title} {The Classical Theory
  of Fields: Volume 2}}},\ Vol.~\bibinfo {volume} {2}\ (\bibinfo  {publisher}
  {Butterworth-Heinemann},\ \bibinfo {year} {1975})\BibitemShut {NoStop}%
\bibitem [{\citenamefont {Zhang}\ and\ \citenamefont
  {Zhang}(2012{\natexlab{a}})}]{zhang2012magnon}%
  \BibitemOpen
  \bibfield  {author} {\bibinfo {author} {\bibfnamefont {S.~S.-L.}\
  \bibnamefont {Zhang}}\ and\ \bibinfo {author} {\bibfnamefont
  {S.}~\bibnamefont {Zhang}},\ }\bibfield  {title} {\bibinfo {title} {Magnon
  mediated electric current drag across a ferromagnetic insulator layer},\
  }\href {https://doi.org/10.1103/PhysRevLett.109.096603} {\bibfield  {journal}
  {\bibinfo  {journal} {Phys. Rev. Lett.}\ }\textbf {\bibinfo {volume} {109}},\
  \bibinfo {pages} {096603} (\bibinfo {year} {2012}{\natexlab{a}})}\BibitemShut
  {NoStop}%
\bibitem [{\citenamefont {Zhang}\ and\ \citenamefont
  {Zhang}(2012{\natexlab{b}})}]{zhang2012spin}%
  \BibitemOpen
  \bibfield  {author} {\bibinfo {author} {\bibfnamefont {S.~S.-L.}\
  \bibnamefont {Zhang}}\ and\ \bibinfo {author} {\bibfnamefont
  {S.}~\bibnamefont {Zhang}},\ }\bibfield  {title} {\bibinfo {title} {Spin
  convertance at magnetic interfaces},\ }\href
  {https://doi.org/10.1103/PhysRevB.86.214424} {\bibfield  {journal} {\bibinfo
  {journal} {Phys. Rev. B}\ }\textbf {\bibinfo {volume} {86}},\ \bibinfo
  {pages} {214424} (\bibinfo {year} {2012}{\natexlab{b}})}\BibitemShut
  {NoStop}%
\bibitem [{\citenamefont {Ritzmann}\ \emph {et~al.}(2014)\citenamefont
  {Ritzmann}, \citenamefont {Hinzke},\ and\ \citenamefont
  {Nowak}}]{ritzmann2014propagation}%
  \BibitemOpen
  \bibfield  {author} {\bibinfo {author} {\bibfnamefont {U.}~\bibnamefont
  {Ritzmann}}, \bibinfo {author} {\bibfnamefont {D.}~\bibnamefont {Hinzke}},\
  and\ \bibinfo {author} {\bibfnamefont {U.}~\bibnamefont {Nowak}},\ }\bibfield
   {title} {\bibinfo {title} {Propagation of thermally induced magnonic spin
  currents},\ }\href {https://doi.org/10.1103/PhysRevB.89.024409} {\bibfield
  {journal} {\bibinfo  {journal} {Phys. Rev. B}\ }\textbf {\bibinfo {volume}
  {89}},\ \bibinfo {pages} {024409} (\bibinfo {year} {2014})}\BibitemShut
  {NoStop}%
\bibitem [{\citenamefont {Cramer}\ \emph {et~al.}(2018)\citenamefont {Cramer},
  \citenamefont {Ritzmann}, \citenamefont {Dong}, \citenamefont {Jaiswal},
  \citenamefont {Qiu}, \citenamefont {Saitoh}, \citenamefont {Nowak},\ and\
  \citenamefont {Kl{\"a}ui}}]{cramer2018spin}%
  \BibitemOpen
  \bibfield  {author} {\bibinfo {author} {\bibfnamefont {J.}~\bibnamefont
  {Cramer}}, \bibinfo {author} {\bibfnamefont {U.}~\bibnamefont {Ritzmann}},
  \bibinfo {author} {\bibfnamefont {B.-W.}\ \bibnamefont {Dong}}, \bibinfo
  {author} {\bibfnamefont {S.}~\bibnamefont {Jaiswal}}, \bibinfo {author}
  {\bibfnamefont {Z.}~\bibnamefont {Qiu}}, \bibinfo {author} {\bibfnamefont
  {E.}~\bibnamefont {Saitoh}}, \bibinfo {author} {\bibfnamefont
  {U.}~\bibnamefont {Nowak}},\ and\ \bibinfo {author} {\bibfnamefont
  {M.}~\bibnamefont {Kl{\"a}ui}},\ }\bibfield  {title} {\bibinfo {title} {Spin
  transport across antiferromagnets induced by the spin {Seebeck} effect},\
  }\href {https://doi.org/10.1088/1361-6463/aab223} {\bibfield  {journal}
  {\bibinfo  {journal} {J. Phys. D: Appl. Phys.}\ }\textbf {\bibinfo {volume}
  {51}},\ \bibinfo {pages} {144004} (\bibinfo {year} {2018})}\BibitemShut
  {NoStop}%
\bibitem [{\citenamefont {Kovalev}\ and\ \citenamefont
  {Zyuzin}(2016)}]{kovalev2016spin}%
  \BibitemOpen
  \bibfield  {author} {\bibinfo {author} {\bibfnamefont {A.~A.}\ \bibnamefont
  {Kovalev}}\ and\ \bibinfo {author} {\bibfnamefont {V.}~\bibnamefont
  {Zyuzin}},\ }\bibfield  {title} {\bibinfo {title} {Spin torque and nernst
  effects in dzyaloshinskii-moriya ferromagnets},\ }\href
  {https://doi.org/10.1103/PhysRevB.93.161106} {\bibfield  {journal} {\bibinfo
  {journal} {Phys. Rev. B}\ }\textbf {\bibinfo {volume} {93}},\ \bibinfo
  {pages} {161106} (\bibinfo {year} {2016})}\BibitemShut {NoStop}%
\bibitem [{\citenamefont {Ivchenko}\ and\ \citenamefont
  {Kiselev}(1973)}]{ivchenko1973spin}%
  \BibitemOpen
  \bibfield  {author} {\bibinfo {author} {\bibfnamefont {E.~L.}\ \bibnamefont
  {Ivchenko}}\ and\ \bibinfo {author} {\bibfnamefont {A.~A.}\ \bibnamefont
  {Kiselev}},\ }\bibfield  {title} {\bibinfo {title} {Spin relaxation of free
  carriers in a noncentrosymmetric semiconductor in a longitudinal magnetic
  field},\ }\href {https://doi.org/?} {\bibfield  {journal} {\bibinfo
  {journal} {Sov. Phys. Solid State}\ }\textbf {\bibinfo {volume} {15}},\
  \bibinfo {pages} {114} (\bibinfo {year} {1973})}\BibitemShut {NoStop}%
\bibitem [{\citenamefont {Glazov}(2004)}]{glazov2004magnetic}%
  \BibitemOpen
  \bibfield  {author} {\bibinfo {author} {\bibfnamefont {M.~M.}\ \bibnamefont
  {Glazov}},\ }\bibfield  {title} {\bibinfo {title} {Magnetic field effects on
  spin relaxation in heterostructures},\ }\href
  {https://doi.org/10.1103/PhysRevB.70.195314} {\bibfield  {journal} {\bibinfo
  {journal} {Phys. Rev. B}\ }\textbf {\bibinfo {volume} {70}},\ \bibinfo
  {pages} {195314} (\bibinfo {year} {2004})}\BibitemShut {NoStop}%
\bibitem [{\citenamefont {Wu}\ \emph {et~al.}(2010)\citenamefont {Wu},
  \citenamefont {Jiang},\ and\ \citenamefont {Weng}}]{wu2010spin}%
  \BibitemOpen
  \bibfield  {author} {\bibinfo {author} {\bibfnamefont {M.}~\bibnamefont
  {Wu}}, \bibinfo {author} {\bibfnamefont {J.}~\bibnamefont {Jiang}},\ and\
  \bibinfo {author} {\bibfnamefont {M.}~\bibnamefont {Weng}},\ }\bibfield
  {title} {\bibinfo {title} {Spin dynamics in semiconductors},\ }\href
  {https://doi.org/10.1016/j.physrep.2010.04.002} {\bibfield  {journal}
  {\bibinfo  {journal} {Phys. Rep.}\ }\textbf {\bibinfo {volume} {493}},\
  \bibinfo {pages} {61} (\bibinfo {year} {2010})}\BibitemShut {NoStop}%
\bibitem [{\citenamefont {Rakitskii}\ \emph {et~al.}(2025)\citenamefont
  {Rakitskii}, \citenamefont {Denisov}, \citenamefont {Rozhansky},\ and\
  \citenamefont {Averkiev}}]{rakitskii2025resonant}%
  \BibitemOpen
  \bibfield  {author} {\bibinfo {author} {\bibfnamefont {M.~A.}\ \bibnamefont
  {Rakitskii}}, \bibinfo {author} {\bibfnamefont {K.~S.}\ \bibnamefont
  {Denisov}}, \bibinfo {author} {\bibfnamefont {I.~V.}\ \bibnamefont
  {Rozhansky}},\ and\ \bibinfo {author} {\bibfnamefont {N.~S.}\ \bibnamefont
  {Averkiev}},\ }\bibfield  {title} {\bibinfo {title} {Resonant spin dynamics
  of {2D} electrons with strong {Rashba} and {Zeeman} couplings},\ }\href
  {https://doi.org/10.1016/j.physe.2024.116147} {\bibfield  {journal} {\bibinfo
   {journal} {Physica E}\ }\textbf {\bibinfo {volume} {166}},\ \bibinfo {pages}
  {116147} (\bibinfo {year} {2025})}\BibitemShut {NoStop}%
\bibitem [{\citenamefont {Denisov}\ and\ \citenamefont
  {{\v{Z}}uti{\'c}}(2024)}]{denisov2024anisotropic}%
  \BibitemOpen
  \bibfield  {author} {\bibinfo {author} {\bibfnamefont {K.~S.}\ \bibnamefont
  {Denisov}}\ and\ \bibinfo {author} {\bibfnamefont {I.}~\bibnamefont
  {{\v{Z}}uti{\'c}}},\ }\bibfield  {title} {\bibinfo {title} {Anisotropic spin
  dynamics in antiferromagnets with a nonrelativistic spin splitting},\ }\href
  {https://doi.org/10.1103/PhysRevB.110.L180403} {\bibfield  {journal}
  {\bibinfo  {journal} {Phys. Rev. B}\ }\textbf {\bibinfo {volume} {110}},\
  \bibinfo {pages} {L180403} (\bibinfo {year} {2024})}\BibitemShut {NoStop}%
\bibitem [{\citenamefont {Su}\ \emph {et~al.}(2024)\citenamefont {Su},
  \citenamefont {Huang}, \citenamefont {Pan},\ and\ \citenamefont
  {Song}}]{su2024controllable}%
  \BibitemOpen
  \bibfield  {author} {\bibinfo {author} {\bibfnamefont {Y.}~\bibnamefont
  {Su}}, \bibinfo {author} {\bibfnamefont {L.}~\bibnamefont {Huang}}, \bibinfo
  {author} {\bibfnamefont {F.}~\bibnamefont {Pan}},\ and\ \bibinfo {author}
  {\bibfnamefont {C.}~\bibnamefont {Song}},\ }\bibfield  {title} {\bibinfo
  {title} {Controllable magnon {Hanle} effect in easy-plane antiferromagnets},\
  }\bibfield  {journal} {\bibinfo  {journal} {Appl. Phys. Lett.}\ }\textbf
  {\bibinfo {volume} {124}},\ \href {https://doi.org/10.1063/5.0210833}
  {10.1063/5.0210833} (\bibinfo {year} {2024})\BibitemShut {NoStop}%
\bibitem [{\citenamefont {Owerre}(2017)}]{owerre2017topological}%
  \BibitemOpen
  \bibfield  {author} {\bibinfo {author} {\bibfnamefont {S.~A.}\ \bibnamefont
  {Owerre}},\ }\bibfield  {title} {\bibinfo {title} {Topological thermal {Hall}
  effect in frustrated {Kagome} antiferromagnets},\ }\href
  {https://doi.org/10.1103/PhysRevB.95.014422} {\bibfield  {journal} {\bibinfo
  {journal} {Phys. Rev. B}\ }\textbf {\bibinfo {volume} {95}},\ \bibinfo
  {pages} {014422} (\bibinfo {year} {2017})}\BibitemShut {NoStop}%
\bibitem [{\citenamefont {Laurell}\ and\ \citenamefont
  {Fiete}(2018)}]{laurell2018magnon}%
  \BibitemOpen
  \bibfield  {author} {\bibinfo {author} {\bibfnamefont {P.}~\bibnamefont
  {Laurell}}\ and\ \bibinfo {author} {\bibfnamefont {G.~A.}\ \bibnamefont
  {Fiete}},\ }\bibfield  {title} {\bibinfo {title} {{Magnon thermal Hall effect
  in kagome antiferromagnets with Dzyaloshinskii-Moriya interactions}},\ }\href
  {https://doi.org/10.1103/PhysRevB.98.094419} {\bibfield  {journal} {\bibinfo
  {journal} {Phys. Rev. B}\ }\textbf {\bibinfo {volume} {98}},\ \bibinfo
  {pages} {094419} (\bibinfo {year} {2018})}\BibitemShut {NoStop}%
\bibitem [{\citenamefont {Doki}\ \emph {et~al.}(2018)\citenamefont {Doki},
  \citenamefont {Akazawa}, \citenamefont {Lee}, \citenamefont {Han},
  \citenamefont {Sugii}, \citenamefont {Shimozawa}, \citenamefont {Kawashima},
  \citenamefont {Oda}, \citenamefont {Yoshida},\ and\ \citenamefont
  {Yamashita}}]{doki2018spin}%
  \BibitemOpen
  \bibfield  {author} {\bibinfo {author} {\bibfnamefont {H.}~\bibnamefont
  {Doki}}, \bibinfo {author} {\bibfnamefont {M.}~\bibnamefont {Akazawa}},
  \bibinfo {author} {\bibfnamefont {H.-Y.}\ \bibnamefont {Lee}}, \bibinfo
  {author} {\bibfnamefont {J.~H.}\ \bibnamefont {Han}}, \bibinfo {author}
  {\bibfnamefont {K.}~\bibnamefont {Sugii}}, \bibinfo {author} {\bibfnamefont
  {M.}~\bibnamefont {Shimozawa}}, \bibinfo {author} {\bibfnamefont
  {N.}~\bibnamefont {Kawashima}}, \bibinfo {author} {\bibfnamefont
  {M.}~\bibnamefont {Oda}}, \bibinfo {author} {\bibfnamefont {H.}~\bibnamefont
  {Yoshida}},\ and\ \bibinfo {author} {\bibfnamefont {M.}~\bibnamefont
  {Yamashita}},\ }\bibfield  {title} {\bibinfo {title} {Spin thermal {Hall}
  conductivity of a kagome antiferromagnet},\ }\href
  {https://doi.org/10.1103/PhysRevLett.121.097203} {\bibfield  {journal}
  {\bibinfo  {journal} {Phys. Rev. Lett.}\ }\textbf {\bibinfo {volume} {121}},\
  \bibinfo {pages} {097203} (\bibinfo {year} {2018})}\BibitemShut {NoStop}%
\bibitem [{\citenamefont {Bae}\ \emph {et~al.}(2022)\citenamefont {Bae},
  \citenamefont {Wang}, \citenamefont {Scheie}, \citenamefont {Xu},
  \citenamefont {Chica}, \citenamefont {Diederich}, \citenamefont {Cenker},
  \citenamefont {Ziebel}, \citenamefont {Bai}, \citenamefont {Ren} \emph
  {et~al.}}]{bae2022exciton}%
  \BibitemOpen
  \bibfield  {author} {\bibinfo {author} {\bibfnamefont {Y.~J.}\ \bibnamefont
  {Bae}}, \bibinfo {author} {\bibfnamefont {J.}~\bibnamefont {Wang}}, \bibinfo
  {author} {\bibfnamefont {A.}~\bibnamefont {Scheie}}, \bibinfo {author}
  {\bibfnamefont {J.}~\bibnamefont {Xu}}, \bibinfo {author} {\bibfnamefont
  {D.~G.}\ \bibnamefont {Chica}}, \bibinfo {author} {\bibfnamefont {G.~M.}\
  \bibnamefont {Diederich}}, \bibinfo {author} {\bibfnamefont {J.}~\bibnamefont
  {Cenker}}, \bibinfo {author} {\bibfnamefont {M.~E.}\ \bibnamefont {Ziebel}},
  \bibinfo {author} {\bibfnamefont {Y.}~\bibnamefont {Bai}}, \bibinfo {author}
  {\bibfnamefont {H.}~\bibnamefont {Ren}}, \emph {et~al.},\ }\bibfield  {title}
  {\bibinfo {title} {Exciton-coupled coherent magnons in a {{2D}}
  semiconductor},\ }\href {https://doi.org/10.1038/s41586-022-05024-1}
  {\bibfield  {journal} {\bibinfo  {journal} {Nature}\ }\textbf {\bibinfo
  {volume} {609}},\ \bibinfo {pages} {282} (\bibinfo {year}
  {2022})}\BibitemShut {NoStop}%
\bibitem [{\citenamefont {Wang}\ \emph {et~al.}(2023)\citenamefont {Wang},
  \citenamefont {Zhang}, \citenamefont {Yang}, \citenamefont {Lin},
  \citenamefont {Chen}, \citenamefont {Yang}, \citenamefont {Gong},
  \citenamefont {Chen}, \citenamefont {Ye},\ and\ \citenamefont
  {Liu}}]{wang2023magnetically}%
  \BibitemOpen
  \bibfield  {author} {\bibinfo {author} {\bibfnamefont {T.}~\bibnamefont
  {Wang}}, \bibinfo {author} {\bibfnamefont {D.}~\bibnamefont {Zhang}},
  \bibinfo {author} {\bibfnamefont {S.}~\bibnamefont {Yang}}, \bibinfo {author}
  {\bibfnamefont {Z.}~\bibnamefont {Lin}}, \bibinfo {author} {\bibfnamefont
  {Q.}~\bibnamefont {Chen}}, \bibinfo {author} {\bibfnamefont {J.}~\bibnamefont
  {Yang}}, \bibinfo {author} {\bibfnamefont {Q.}~\bibnamefont {Gong}}, \bibinfo
  {author} {\bibfnamefont {Z.}~\bibnamefont {Chen}}, \bibinfo {author}
  {\bibfnamefont {Y.}~\bibnamefont {Ye}},\ and\ \bibinfo {author}
  {\bibfnamefont {W.}~\bibnamefont {Liu}},\ }\bibfield  {title} {\bibinfo
  {title} {Magnetically-dressed {CrSBr} exciton-polaritons in ultrastrong
  coupling regime},\ }\href {https://doi.org/10.1038/s41467-023-41688-7}
  {\bibfield  {journal} {\bibinfo  {journal} {Natur. Commun.}\ }\textbf
  {\bibinfo {volume} {14}},\ \bibinfo {pages} {5966} (\bibinfo {year}
  {2023})}\BibitemShut {NoStop}%
\bibitem [{\citenamefont {Liang}\ \emph {et~al.}()\citenamefont {Liang},
  \citenamefont {Yan}, \citenamefont {Hao}, \citenamefont {Zhu}, \citenamefont
  {Sui}, \citenamefont {Pharis}, \citenamefont {Huang}, \citenamefont {Jain},
  \citenamefont {Zhou}, \citenamefont {Yi} \emph
  {et~al.}}]{liang2026ferroelectric}%
  \BibitemOpen
  \bibfield  {author} {\bibinfo {author} {\bibfnamefont {Y.}~\bibnamefont
  {Liang}}, \bibinfo {author} {\bibfnamefont {X.}~\bibnamefont {Yan}}, \bibinfo
  {author} {\bibfnamefont {B.}~\bibnamefont {Hao}}, \bibinfo {author}
  {\bibfnamefont {Z.}~\bibnamefont {Zhu}}, \bibinfo {author} {\bibfnamefont
  {T.}~\bibnamefont {Sui}}, \bibinfo {author} {\bibfnamefont {D.}~\bibnamefont
  {Pharis}}, \bibinfo {author} {\bibfnamefont {X.}~\bibnamefont {Huang}},
  \bibinfo {author} {\bibfnamefont {R.}~\bibnamefont {Jain}}, \bibinfo {author}
  {\bibfnamefont {T.}~\bibnamefont {Zhou}}, \bibinfo {author} {\bibfnamefont
  {D.}~\bibnamefont {Yi}}, \emph {et~al.},\ }\bibfield  {title} {\bibinfo
  {title} {Ferroelectric switching of odd-parity magnon spin splitting},\
  }\href@noop {} {\bibinfo  {journal} {arXiv:2609.24182}\ }\BibitemShut
  {NoStop}%
\end{thebibliography}%
	
\end{document}